\documentclass[preprintnumbers,amsmath,amssymb,floatfix,10pt,prd,twocolumn,
superscriptaddress,nofootinbib]{revtex4-2}
\usepackage{latexsym}
\usepackage{epsfig}
\usepackage{epstopdf}
\usepackage{graphicx}
\usepackage{amssymb}
\usepackage{amsmath}
\usepackage{amsfonts}
\usepackage{subfigure}
\usepackage{dcolumn}
\usepackage{bm}
\usepackage[normalem]{ulem}
\usepackage[dvipsnames]{xcolor}
\usepackage{hyperref}
\usepackage{color}
\usepackage{comment}
\usepackage{float}
\usepackage[utf8]{inputenc}
\hypersetup{colorlinks=true,linkcolor=blue,citecolor=red,urlcolor=magenta}
\usepackage{mathrsfs}
\usepackage{mathtools}
\usepackage{orcidlink}
\usepackage{appendix}
\usepackage{booktabs}
\newcommand{\dd}{\mathrm{d}}
\newcommand{\e}{\mathrm{e}}

\begin{document}

\title{Black holes from a nonlocal equation of state: horizon geometry, singularity structure and thermodynamics}

\author{Hishan Farfán-Bachiloglu \orcidlink{0000-0002-1382-5605}}
\email{hishan.farfan@ucn.cl}
\affiliation{Departamento de F\'isica, Universidad Cat\'olica del Norte, Av. Angamos 0610, Antofagasta, Chile.}

\author{C. R. Muniz \orcidlink{0000-0002-1266-2218}}
\email{celio.muniz@uece.br}
\affiliation{Universidade Estadual do Cear\'a, Faculdade de Educa\c c\~ao, Ci\^encias e Letras de Iguatu, 63500-000, Iguatu, CE, Brazil.}

\author{Francisco Tello-Ortiz \orcidlink{0000-0002-7104-5746}}
\email{francisco.tello@ufrontera.cl}
\affiliation{Departamento de Ciencias F\'isicas, Universidad de La Frontera, Casilla 54-D, 4811186 Temuco, Chile.}

\begin{abstract}
We show that, for a static and spherically symmetric geometry of Kerr--Schild type sourced by a single anisotropic fluid, a nonlocal equation of state in which the radial pressure depends on the matter enclosed within each radius selects the density profile uniquely: the closure reduces to a linear equation for the enclosed matter, and its solution is a finite-range hair of Yukawa type, exponentially screened at large radius, with the asymptotic mass as the only conserved charge. The resulting neutral family admits an exact classification of its horizons: depending on the amplitude and the range of the hair it describes horizonless configurations, Schwarzschild-like black holes, and black holes with an inner Cauchy horizon, separated by an extremal curve that we obtain in closed form together with the mass and radius of the zero-temperature extremal configuration. We prove that the inner surface gravity exceeds the outer one throughout the two-horizon region. When the central mass vanishes, the center is a global monopole where it is static and a Kantowski--Sachs-type region where it is not, and the singularity satisfies Kr\'olak's strong-curvature condition, with a Tipler-strong branch along non-radial geodesics. In this case, the curvature magnitudes are also locally integrable with respect to the invariant spacetime volume. Replacing the volume weight of the closure by a power-law weight produces a family of solutions in terms of incomplete gamma functions that interpolates between this singular core and a regular black hole with a de~Sitter core, obtained from the same constitutive principle without nonlinear electrodynamics. The surface gravities, the extended first law and the Smarr relation take simple closed forms, and the heat capacity at fixed hair changes sign at a point determined by the lower branch of the Lambert function. A thermodynamic topological analysis identifies two distinct classes governed by the gravitational hair: a Schwarzschild-like regime with a single thermodynamic defect and an extremal regime with two defects of opposite winding numbers, which merge at the maximum Hawking temperature.
\end{abstract}

\maketitle

\section{Introduction}
\label{sec:intro}

In General Relativity the end state of gravitational collapse is described by a remarkably small family of solutions. Stationary black holes of the Einstein--Maxwell system are fixed by mass, angular momentum and charge \cite{Israel:1967wq,Carter:1971zc,Robinson:1975bv,Kerr:1963ud,Newman:1965my,Chrusciel:2012jk}, a rigidity that underlies the no-hair conjecture \cite{Ruffini:1971bza,Hawking:1971vc} and the scalar no-hair theorems \cite{Bekenstein:1995un,Mayo:1996mv}. Within this class the causal structure follows a rigid pattern. A static neutral black hole has a single horizon and a spacelike singularity, whereas an inner, Cauchy horizon and a timelike singularity appear only when charge or rotation is added. The inner horizon is not a harmless feature: it marks the boundary of predictability, it is the locus of the blueshift instability and of mass inflation \cite{Penrose:1969pc,Poisson:1989zz,Poisson:1990eh,Ori:1991zz}, and its fate in the presence of matter fields and of nonlinear dynamics is an active subject \cite{German:2001ti,Dafermos:2003vim,Hartnoll:2020rwq,Cai:2020wrp,An:2021plu,Devecioglu:2021xug,Carballo-Rubio:2018pmi,Ovalle:2023vvu}.

Outside the Einstein--Maxwell class, matter alone can give a static black hole an inner horizon: regular black holes with a de~Sitter core and several anisotropic-fluid geometries already have this property. The question we address is different and concerns the origin of the profile: can a constitutive relation of the matter, rather than a prescribed metric, select a neutral family whose horizon structure, including the two-horizon region and the extremal curve, is analytically controlled, and what does the resulting geometry look like near the origin and at the horizons? Hair is known to arise in controlled ways. Black holes immersed in matter, the so-called dirty black holes, keep the basic structure of their horizons while the exterior is reshaped by the surrounding distribution \cite{Visser:1992qh}. Scalar fields with suitable couplings support hairy solutions that evade the classical theorems \cite{Martinez:2002ru,Martinez:2004nb,Herdeiro:2015waa}, and halos or quintessence-like fluids provide analytic examples in which a matter sector modifies the strong-field metric while the asymptotic charges remain those of the vacuum solution \cite{Kiselev:2002dx,Konoplya:2019sns,Konoplya:2022hbl,Heydarzade:2023dof}. Regular black holes, in which the central singularity is replaced by a core, show that matter with a suitable profile can also change the interior qualitatively \cite{Dymnikova:1992ux,AyonBeato:1998ub,Hayward:2005gi,Bonanno:2000ep}. What is usually missing in these constructions is a principle that selects the matter profile.

The difficulty is that in a large fraction of the literature the metric is prescribed first, frequently by inserting an exponential, Yukawa-type or power-law correction into a known solution, and the source is reconstructed afterwards. The matter content then has to be examined a posteriori: whether the energy density has a definite sign, whether the standard energy conditions hold \cite{Hawking:1973uf,Visser:1997tq,Curiel:2014zba}, and how these requirements constrain the number and nature of the horizons \cite{Yang:2021civ}. Yukawa-type black holes have indeed been obtained along several distinct routes. In higher-derivative gravity the massive spin-two mode produces Yukawa terms in the static field \cite{Stelle:1977ry,Stelle:1977gr}, and the corresponding black-hole solutions have been studied in detail \cite{Lu:2015cqa,Lu:2015psa}; exponentially suppressed, Yukawa-like corrections also arise in nonlocal Deser--Woodard gravity \cite{Li:2026drw}. A third route starts from a Yukawa-modified gravitational potential, converts it into an effective energy density, and integrates the Einstein equations, with or without electric charge \cite{Gonzalez:2023rsd,AraujoFilho:2023ibe,AraujoFilho:2024ctw}. In each case the Yukawa profile is inherited from an assumed potential or from the extra gravitational degrees of freedom of the underlying theory, rather than from a property of the matter that sources the geometry. The constructions available in the literature can thus be grouped as (i) metrics in which the Yukawa factor is inserted by hand, (ii) metrics integrated from an effective density built on a Yukawa-modified potential \cite{Gonzalez:2023rsd,AraujoFilho:2023ibe,AraujoFilho:2024ctw}, and (iii) solutions of modified gravity in which the factor is carried by additional gravitational modes \cite{Stelle:1977ry,Stelle:1977gr,Lu:2015cqa,Lu:2015psa,Li:2026drw}. The present construction belongs to a fourth class: (iv) solutions of General Relativity in which the profile is the unique output of an explicit constitutive relation of the source. A framework in which the profile of the matter sector follows from a constitutive principle within General Relativity, rather than from a choice of the metric or of the potential, is therefore highly desirable. Finite-range, Yukawa-type behavior is a particularly well-motivated target: it is ubiquitous in massive and effective descriptions of gravity and matter, it preserves the asymptotic charges, and it provides the simplest setting in which hair is visible in the strong-field region while screened at large distances.

Anisotropic matter offers the natural language for such a construction. Local anisotropy allows the radial and tangential stresses to respond differently to the geometry, and nonlocal equations of state go one step further by letting the pressure at a given radius depend on the matter enclosed within it. Nonlocal closures were introduced in the study of relativistic fluid spheres \cite{Hernandez:1998hk,Hernandez:2001nr} and encode a collective property of the distribution that a purely local equation of state cannot capture. For static spherical symmetry, the Kerr--Schild condition \cite{Kerr:1965wfc,Gurses:1975vu,Jacobson:2007tj} ties the radial pressure to the density and therefore constrains the admissible matter very strongly. Putting the two ingredients together turns the question above into a well-posed one: which radial profile is compatible with both a nonlocal closure and the Kerr--Schild structure, with no other field present, and what geometry does it generate? 
The answer is an exact solution in which the Yukawa profile is a consequence of the closure and not an input. The nonlocal closure selects a neutral family with an analytic two-horizon region and an analytic extremal curve: besides the mass, the amplitude and the range of the hair generate black holes with a single horizon, black holes with a Cauchy horizon, extremal configurations and configurations without a horizon, and the extremal curve and the zero-temperature extremal configuration can be written in closed form. The inner surface gravity is shown to exceed the outer one in the whole two-horizon region. Where the central region is static it is that of a global monopole \cite{Barriola:1989hx}, and we test the strength of the singularity with the criteria of Clarke, Kr\'olak and Tipler \cite{Tipler:1980,Clarke:1985jg,Krolak:1987}. The thermodynamics, formulated with the amplitude and the range of the hair as additional variables, extends the laws of black-hole mechanics \cite{Bekenstein:1973ur,Bardeen:1973gs,Hawking:1975vcx,Smarr:1973zz,Iyer:1994ys,Davies:1977bgr} along the lines of recent extended formulations \cite{Xiao:2023lap,Hajian:2023cwe}. Finally, the nonlocal closure admits a one-parameter family of weights, and a member of it produces a regular black hole with a de~Sitter core, so that the same constitutive scheme connects a singular and a regular core.

The article is organized as follows. Sect.~\ref{sec:sol} derives the solution and describes its matter content and curvature. Sect.~\ref{sec:hor} classifies the horizon structure and obtains the zero-temperature extremal configuration. Sect.~\ref{sec:core} analyzes the central region, the strength of the singularity, and the local integrability of curvature invariants. Sect.~\ref{sec:thermo} develops the extended and topological thermodynamics, as well as the properties of the inner horizon. Sect.~\ref{sec:family} introduces the family of closures and its regular member. Sect.~\ref{sec:concl} summarizes the results. We use the signature $(-,+,+,+)$ and geometrized units $G=c=1$.

%=====================================================================
\section{Field equations, nonlocal closure and solution}
\label{sec:sol}

\subsection{Kerr--Schild geometry and the Einstein equations}

We consider a static and spherically symmetric geometry in Schwarzschild gauge,
\begin{equation}
\dd s^2=-f(r)\,\dd t^2+\frac{\dd r^2}{f(r)}+r^2\dd\Omega^2,\quad \dd\Omega^2=\dd\theta^2+\sin^2\theta\,\dd\phi^2,
\label{metric}
\end{equation}
with $f(r)=1-2m(r)/r$.
Passing to the advanced time $v=t+\int\dd r/f$ and then to $\tilde t=v-r$, the line element takes the Kerr--Schild form
\begin{equation}
g_{\mu\nu}=\eta_{\mu\nu}+\frac{2m(r)}{r}\,k_\mu k_\nu,\qquad k_\mu\dd x^\mu=\dd\tilde t+\dd r,
\label{KS}
\end{equation}
where $\eta_{\mu\nu}$ is the Minkowski metric and $k^\mu$ is a null vector tangent to a geodesic, shear-free and twist-free congruence with expansion $2/r$. The principal null directions of the Weyl tensor are the radial null directions, so that the geometry is of Petrov type~D, with
\begin{equation}
\Psi_2=-\frac{m}{r^3}+\frac{2m'}{3r^2}-\frac{m''}{6r}.
\label{psi2}
\end{equation}

The only matter content is an anisotropic fluid with stress tensor $T^\mu{}_\nu=\mathrm{diag}(-\rho,p_r,p_t,p_t)$. The $tt$, $rr$ and $\theta\theta$ components of the Einstein equations $G^\mu{}_\nu=8\pi T^\mu{}_\nu$ give
\begin{equation}
m'=4\pi r^2\rho,\qquad p_r=-\rho,\qquad p_t=-\frac{m''}{8\pi r}.
\label{einstein}
\end{equation}
The second relation is the geometric content of the Kerr--Schild structure: it is equivalent to $R_{\mu\nu}k^\mu k^\nu=8\pi(\rho+p_r)=0$, that is, to the statement that the null congruence generating the geometry is Ricci-flat along itself, and it follows from the one-function ansatz, which forces $T^t{}_t=T^r{}_r$ \cite{Jacobson:2007tj}. The radial conservation equation $p_r'=-2(p_r-p_t)/r$ is a consequence of Eq.~\eqref{einstein}. The configuration is thus a ``dirty'' black hole in the sense of Ref.~\cite{Visser:1992qh}, and the whole problem reduces to the determination of a single radial profile $\rho(r)$.

\subsection{The nonlocal equation of state}

We close the system with a nonlocal equation of state, in which the radial pressure at $r$ depends on the matter enclosed within that radius,
\begin{equation}
p_r=\rho+\frac{A}{r^2}\int_0^r x^2\rho(x)\,\dd x+\frac{\sigma}{r^2}.
\label{nleos}
\end{equation}
Here $A$ has dimensions of inverse length and sets the range of the nonlocality, and $\sigma$ is a constant. Each term of Eq.~\eqref{nleos} has a direct interpretation. The weight $x^2$ is the volume measure of the spherical shells, so that by Eq.~\eqref{einstein}
\begin{equation}
\frac{A}{r^2}\int_0^r x^2\rho\,\dd x=\frac{A}{4\pi}\,\frac{m(r)-m(0)}{r^2},
\end{equation}
and the nonlocal term is the matter mass enclosed by the sphere of radius $r$ divided by its area, the same combination that sets the Newtonian field at $r$. The closure therefore makes the radial stress respond to the enclosed matter with a single constant coupling $A$, which is the simplest choice introducing one length scale and keeping the relation linear in $\rho$. The term $\sigma/r^2$ is the homogeneous part of the relation and is the only seed of the matter sector: with $\mathcal{Q}(0)=0$ and $\sigma=0$ the equation below gives $\rho\equiv0$, and the geometry is Schwarzschild. Near the center $\mathcal{Q}\to0$, and $\sigma/r^2$ is balanced by $2\rho$, which fixes the $r^{-2}$ behavior of the density; it is not cancelled by the solution, and it is the constitutive origin of the monopole-like core of Sect.~\ref{sec:core}. The dependence on the enclosed matter is a radial relation within a static configuration; it is a constitutive closure of the static problem, not a causal law of evolution for the fluid. Introducing the enclosed quantity $\mathcal{Q}(r)=\int_0^r x^2\rho\,\dd x$, with $\mathcal{Q}(0)=0$ and $\mathcal{Q}'=r^2\rho$, and using $p_r=-\rho$, Eq.~\eqref{nleos} becomes the linear first-order equation
\begin{equation}
2\mathcal{Q}'+A\,\mathcal{Q}+\sigma=0,
\label{Qode}
\end{equation}
whose solution is $\mathcal{Q}(r)=-(\sigma/A)\,(1-\e^{-Ar/2})$. Therefore
\begin{equation}
\rho(r)=-\frac{\sigma}{2r^2}\,\e^{-Ar/2}=-\frac{A\,l}{16\pi r^2}\,\e^{-Ar/2},\qquad l\equiv\frac{8\pi\sigma}{A}.
\label{rho}
\end{equation}
The only constant that survives is $\sigma$, equivalently the amplitude $l$: the closure fixes both the shape of the profile and its range. Integrating $m'=4\pi r^2\rho$ with the asymptotic condition $m(\infty)=M$ gives
\begin{equation}
m(r)=M+\frac{l}{2}\,\e^{-Ar/2},\qquad f(r)=1-\frac{2M}{r}-\frac{l}{r}\,\e^{-Ar/2},
\label{sol}
\end{equation}
and the tangential pressure follows from Eq.~\eqref{einstein},
\begin{equation}
p_t=\frac{Ar}{4}\,\rho=-\frac{A^2 l}{64\pi r}\,\e^{-Ar/2}.
\label{pt}
\end{equation}
The geometry is an exact solution of the Einstein equations sourced by one anisotropic fluid. The parameter $M$ is the asymptotic mass and is the only conserved charge: the hair is screened as $\e^{-Ar/2}$, so the range of the Yukawa sector is $2/A$. For $l<0$ the fluid carries a positive total mass $-l/2=4\pi\int_0^\infty r^2\rho\,\dd r$, and the central mass is $m(0)=M+l/2=M-|l|/2$. The limit $A\to0$ depends on which constitutive constant is held fixed. At fixed amplitude $l$ the density vanishes, $\sigma=Al/8\pi\to0$, and the geometry becomes Schwarzschild with mass $M+l/2$. At fixed $\sigma$ the amplitude $l=8\pi\sigma/A$ diverges, and the limit is regular only if the central mass $m_0=M+4\pi\sigma/A$ is kept fixed; then $m(r)\to m_0-2\pi\sigma r$ and
\begin{equation}
f(r)\to1-4\pi|\sigma|-\frac{2m_0}{r},\qquad \sigma<0,
\end{equation}
which is the Barriola--Vilenkin black hole with a global monopole of solid-angle deficit $4\pi|\sigma|$ \cite{Barriola:1989hx}, the density $\rho=|\sigma|/(2r^2)$ being unscreened. The two limits are physically distinct: the first switches the fluid off, the second removes its screening. For $l\to0$ the geometry reduces to Schwarzschild with mass $M$.

Interestingly, the density profile obtained in Eq.~(\ref{rho}) coincides with a particular generalized Einasto distribution \cite{Fielder:2020lpk},
\begin{equation}
\rho_{\mathrm{gE}}(r)=\rho_s\left(\frac{r}{r_s}\right)^{-\gamma}
\exp\!\left[-\frac{2}{\alpha}\left(\left(\frac{r}{r_s}\right)^\alpha-1\right)\right].
\end{equation}
Choosing $\alpha=1$, $\gamma=2$, $r_s=4/A$, and $\rho_s=-A^3l/(256\pi e^2)$ exactly recovers Eq.~(\ref{rho})
for the positive-density branch $l<0$. Thus, nonlocal closure~(\ref{nleos}), together with the Kerr–Schild condition, selects a density profile equivalent to a singular isothermal profile with exponential cutoff, belonging to the generalized Einasto family. The selected parameters differ from those typically found in dark matter halo simulations, and this similarity does not imply equivalent underlying dynamics. The central behavior of the density also gives rise to the global-monopole geometry discussed in Sec. \ref{sec:core} when the central Schwarzschild contribution vanishes.

\subsection{Matter content and curvature}

For $l<0$ the energy density is positive and the ratio $p_t/\rho=Ar/4$ is positive. The null, weak and strong energy conditions then hold at every radius, and the dominant energy condition holds if and only if $Ar\le4$. The scalar curvature, the Kretschmann scalar and the Weyl scalar are
\begin{align}
R&=\frac{4m'}{r^2}+\frac{2m''}{r}=-\frac{A\,l}{r^2}\left(1-\frac{Ar}{4}\right)\e^{-Ar/2},\label{Ricci}\\
K&=f''^2+\frac{4f'^2}{r^2}+\frac{4(1-f)^2}{r^4},
\end{align}
with $\Psi_2$ given by Eq.~\eqref{psi2}. The Ricci scalar and the matter invariants decay exponentially at large radius, while the Kretschmann scalar approaches its Schwarzschild value, so the hair is a strong-field feature.

\paragraph{Parameters and scope of the closure.} The constant $A>0$ has dimensions of inverse length and is the inverse of half the range of the hair; $A<0$ would produce an exponentially growing profile and is excluded. The amplitude $l$ has dimensions of length, and $\sigma=Al/8\pi$ is dimensionless; $l<0$ (equivalently $\sigma<0$) corresponds to a positive density, while $l>0$ gives a negative density and is kept only as a mathematical branch. The fluid is an effective description: the closure~\eqref{nleos} is postulated, as a relation that makes the pressure at a radius depend on the matter enclosed within it, and not derived from a microscopic model. It is useful to separate what follows from the form of $f(r)$ in Eq.~\eqref{sol} alone from what requires the closure. The horizon structure, the surface gravities, the temperature and the first law follow from $f(r)$ with $M$, $l$ and $A$ as free parameters, whatever the origin of the profile. The closure is what selects this profile, ties the amplitude of the density to the constant $\sigma$, fixes the ratio $p_t/\rho=Ar/4$ and with it the energy conditions, and generates the family of Sect.~\ref{sec:family}. At an inner horizon, $f'(r_-)\le0$ translates through Eq.~\eqref{einstein} into $8\pi r_-^2\rho(r_-)\ge1$, so an inner horizon requires a density that is large enough there; no energy condition forbids this, since $\rho>0$ and $\rho+p_i\ge0$ hold.

%=====================================================================
\section{Horizon structure and the extremal configuration}
\label{sec:hor}

The classification is expressed in terms of the dimensionless combinations collected in Table~\ref{tab:param}, which are used throughout the rest of the paper.

\begin{table}[h]
\centering\small
\caption{Dimensionless parameters of the solution.}
\label{tab:param}
\begin{tabular}{ll}
\toprule
Combination & Meaning\\
\midrule
$AM$ & mass in units of the inverse range\\
$l/M$ & hair amplitude relative to the mass\\
$\mathcal{K}=|l|A/2$ ($l<0$) & strength of the hair at fixed $(l,A)$\\
$\lambda=lA=-2\mathcal{K}$ & signed form of $\mathcal{K}$ used in Fig.~\ref{fig:thermo}\\
$x=Ar/2$ & radius in units of the range $2/A$\\
\bottomrule
\end{tabular}
\end{table}

Writing the metric function as
\begin{equation}
f(r)=\frac{\e^{-Ar/2}}{r}\,\bigl[g(r)-l\bigr],\qquad g(r)\equiv(r-2M)\,\e^{Ar/2},
\label{fg}
\end{equation}
the horizons are the positive roots of $g(r)=l$. The physical domain is $r>0$: a root at $r=0$ is not a horizon, since $r=0$ is a curvature singularity for every $l\neq0$ (Sect.~\ref{sec:core}). The function $g$ has $g(0)=-2M$ and $g'(r)=\e^{Ar/2}\bigl[1+A(r-2M)/2\bigr]$.

\textbf{Proposition 1.}
\textit{
Let $M>0$.
\begin{enumerate}
\item If $AM\le1$, the metric has exactly one horizon when $l>-2M$ and none when $l\le-2M$.
\item If $AM>1$, define $r_\ast=2M-2/A>0$ and $l_c(A)=-\frac{2}{A}\,\e^{AM-1}<-2M$. The metric has one horizon for $l\ge-2M$, two horizons $r_-<r_+$ for $l_c<l<-2M$, a single degenerate horizon at $r_\ast$ for $l=l_c$, and none for $l<l_c$.
\end{enumerate}
}

\textit{Proof.} 
If $AM\le1$ then $g'>0$ for all $r>0$ (at the boundary value $AM=1$ one has $g'(0)=0$, but $g'>0$ for $r>0$, so the marginal case belongs to the monotonic class), so $g$ increases monotonically from $-2M$ to infinity and the equation $g=l$ has one solution for $l>-2M$ and none otherwise. If $AM>1$, $g'<0$ on $(0,r_\ast)$ and $g'>0$ on $(r_\ast,\infty)$, so $g$ decreases from $-2M$ to its minimum $g(r_\ast)=l_c$ and then increases without bound. Each branch is monotonic and therefore contains at most one root. For $l>-2M$ only the increasing branch is crossed; for $l=-2M$ the root at $r=0$ is not a horizon and a single root remains on the increasing branch; for $l_c<l<-2M$ both branches are crossed; for $l=l_c$ the two roots merge at $r_\ast$, where $f=f'=0$. The inequality $l_c<-2M$ follows from $\e^{y-1}>y$ for $y\neq1$.

\begin{figure}[t]
\centering
\includegraphics[width=\columnwidth]{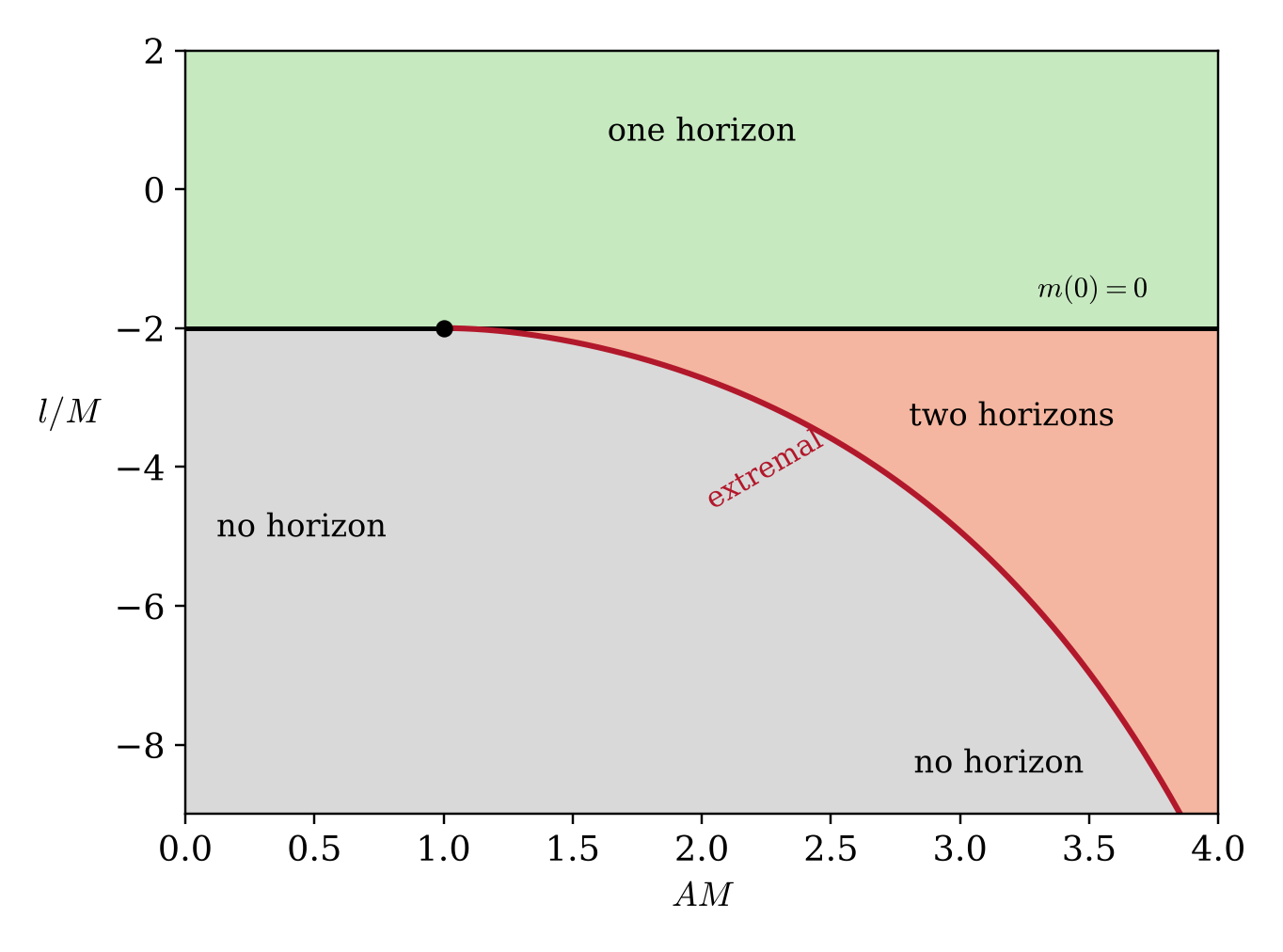}
\caption{Classification of the geometry in the plane of the hair range $AM$ and amplitude $l/M$. The curve $l=-2M$, where the central mass vanishes, and the extremal curve $l=l_c(A)$ meet at $AM=1$. Between them the geometry has two horizons.}
\label{fig:map}
\end{figure}

\begin{figure}[t]
\centering
\includegraphics[width=\columnwidth]{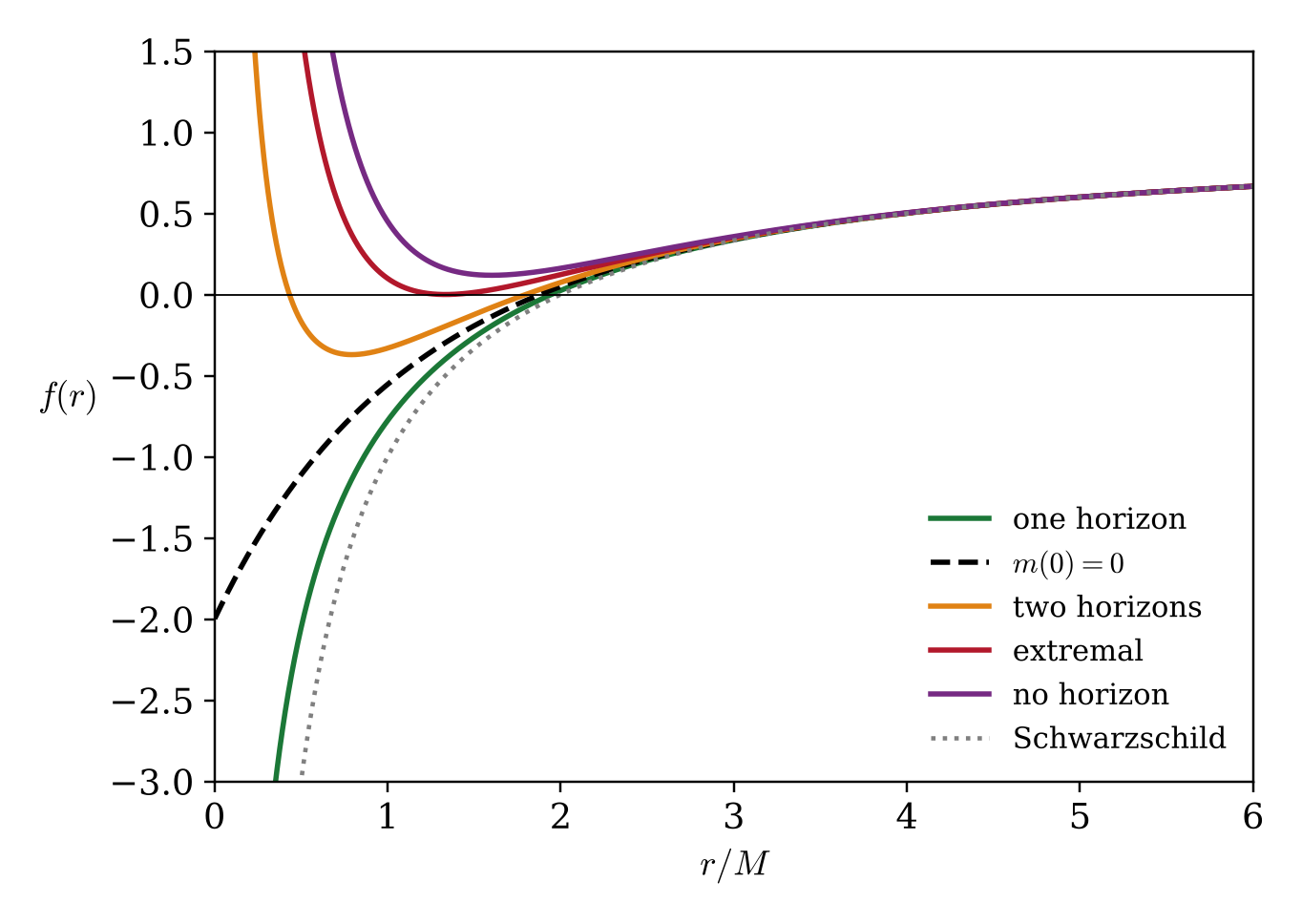}
\caption{Metric function for representative values of the hair amplitude at fixed range: one horizon, vanishing central mass, two horizons, extremal and no horizon. The Schwarzschild profile is shown for comparison.}
\label{fig:prof}
\end{figure}

At most two horizons exist, as for Reissner--Nordstr\"om, although the horizon equation is transcendental. The classification is shown in Fig.~\ref{fig:map}, and the profiles of the metric function in the different regimes are shown in Fig.~\ref{fig:prof}.

\subsection{Zero-temperature extremal configuration in closed form}

It is more useful to read the classification at fixed hair, that is, at fixed $(l,A)$, as a statement about the mass. Let $\mathcal{K}\equiv|l|A/2$ for $l<0$. If $\mathcal{K}\le1$, a horizon exists if and only if $M>|l|/2$. If $\mathcal{K}>1$, the degenerate horizon sits at
\begin{equation}
r_{\rm ext}=\frac{2}{A}\ln\mathcal{K},\qquad M_{\rm ext}=\frac{1+\ln\mathcal{K}}{A},
\label{ext}
\end{equation}
which follow from $l\,\e^{-Ar_{\rm ext}/2}=-2/A$ and $f(r_{\rm ext})=0$. A horizon exists if and only if $M\ge M_{\rm ext}$; for $M_{\rm ext}<M<|l|/2$ there are two horizons, for $M=|l|/2$ the central mass vanishes, and for $M>|l|/2$ there is a single horizon. Since $1+\ln\mathcal{K}<\mathcal{K}$ for $\mathcal{K}\ne1$, one has $M_{\rm ext}<|l|/2$, and the two expressions coincide at $\mathcal{K}=1$, where the inner horizon disappears at the origin. Thus the static family contains a neutral configuration with a finite minimal mass and a degenerate horizon, a property usually associated with charge, and Eq.~\eqref{ext} gives it in closed form. Section~\ref{sec:thermo} discusses it as a candidate end point of evaporation.

\subsection{Causal structure}

The Penrose diagrams follow from the classification. In the regime of a single horizon with $M>|l|/2$ (or $l\ge0$) the central mass is positive, $f\to-\infty$ at the origin, and the singularity is spacelike as in Schwarzschild. When two horizons are present the central mass is negative, $f\to+\infty$ at the origin, the singularity is timelike and repulsive, and the inner horizon is a Cauchy horizon, as in Reissner--Nordstr\"om. At $l_c$ the horizons merge into a degenerate one. Below $l_c$, and for $l\le-2M$ when $AM\le1$, the singularity is naked. The case $M=|l|/2$ is discussed in the next section.

%=====================================================================
\section{Central region, strength of the singularity, and curvature integrability}
\label{sec:core}

Expanding the mass function near the origin, $m(r)=m_0+m_1r+\tfrac12m_2r^2+\dots$, one finds from Eq.~\eqref{sol}
\begin{equation}
m_0=M+\frac l2,\qquad m_1=-\frac{Al}{4},\qquad m_2=\frac{A^2l}{8}.
\end{equation}
When $m_0\ne0$ the leading behavior is $f\simeq-2m_0/r$, the Kretschmann scalar diverges as $48\,m_0^2/r^6$ and the singularity is of Schwarzschild type for $m_0>0$ and of Reissner--Nordstr\"om type for $m_0<0$.

The interesting case is $m_0=0$, that is $l=-2M$, in which the Schwarzschild term is exactly cancelled by the hair. The metric function is then finite at the origin,
\begin{equation}
f(r)=f_0+\frac{A^2|l|}{8}\,r+\mathcal{O}(r^2),\qquad f_0=1-\frac{A|l|}{2}=1-AM,
\label{f0}
\end{equation}
but the geometry is not regular. The density behaves as $\rho\simeq A|l|/(16\pi r^2)$ with $p_r=-\rho$, while $p_t$ is suppressed by a power of $r$, and the curvature invariants are
\begin{equation}
R\simeq\frac{A|l|}{r^2},\qquad K\simeq\frac{4(1-f_0)^2}{r^4}=\frac{A^2l^2}{r^4},
\end{equation}
for either sign of $f_0$, so the singularity is a curvature singularity, much milder than the $r^{-6}$ divergence of Schwarzschild. The causal character of the central region depends on the sign of $f_0$, and the geometric interpretation differs.

\paragraph{The branch $f_0>0$ ($AM<1$).} The rescaling $t'=\sqrt{f_0}\,t$, $r'=r/\sqrt{f_0}$ turns the leading metric into
\begin{equation}
\dd s^2\simeq-\dd t'^2+\dd r'^2+f_0\,r'^2\,\dd\Omega^2,
\end{equation}
the metric of a global monopole \cite{Barriola:1989hx} with solid-angle deficit $1-f_0=A|l|/2$. The hair therefore realizes a global monopole whose strength is set by the product of its amplitude and its inverse range; the singularity is timelike, and the density and radial pressure are those of the monopole.

\paragraph{The branch $f_0<0$ ($AM>1$).} Here the region $r<r_+$ is a trapped region in which $r$ is a time coordinate, and the static interpretation does not apply. With $\tau=r/\sqrt{|f_0|}$ and $x=\sqrt{|f_0|}\,t$ the leading metric is
\begin{equation}
\dd s^2\simeq-\dd\tau^2+\dd x^2+|f_0|\,\tau^2\,\dd\Omega^2,
\end{equation}
a homogeneous Kantowski--Sachs-type geometry in which the spheres grow linearly with $\tau$. The singularity is spacelike, the invariants are those given above, and the deficit-angle language of the monopole is not used. The marginal case $f_0=0$ ($AM=1$) has $f\propto r$ at the origin and the same leading invariants.

\paragraph{Strength of the singularity.} We use the criteria of Tipler and of Kr\'olak \cite{Tipler:1980,Clarke:1985jg,Krolak:1987}, which give sufficient conditions in terms of the Ricci contraction along timelike geodesics. A singularity is strong in the sense of Tipler (Kr\'olak) along a geodesic if every volume element spanned by Jacobi fields along it is crushed to zero (decreases monotonically) as the singularity is reached. The integral conditions below are the sufficient conditions of Refs.~\cite{Clarke:1985jg,Krolak:1987}, which replace the analysis of the Jacobi fields by the divergence of the integrated Ricci contraction; each statement below applies to the class of geodesics on which it is evaluated, which we specify explicitly together with its accessibility to the center. For a timelike geodesic with energy $E$ and angular momentum $L$ per unit mass, in the equatorial plane, one finds from the field equations
\begin{equation}
\begin{split}R_{\mu\nu}u^\mu u^\nu=&\,8\pi\rho\,\frac{L^2}{r^2}+8\pi p_t\left(1+\frac{L^2}{r^2}\right).\end{split}
\end{equation}
For radial geodesics ($L=0$) this reduces to $8\pi p_t=(A^2|l|/8r)\,\e^{-Ar/2}$. A radial geodesic that reaches $r=0$ has $\dot r^2=E^2-f\to E^2-f_0>0$ (for $f_0<0$ always, and for $f_0>0$ when it comes from infinity with $E\ge1$), so $r\simeq c\,\tau$ in terms of the proper time $\tau$ remaining to the singularity, with $c=\sqrt{E^2-f_0}$, and $R_{\mu\nu}u^\mu u^\nu\simeq A^2|l|/(8c\,\tau)$. Then
\begin{align}
\int_\tau^{\tau_0}R_{\mu\nu}u^\mu u^\nu\,\dd s&=\frac{A^2|l|}{8c}\ln\frac{\tau_0}{\tau}\to\infty,\\
\int_\tau^{\tau_0}\!\dd s\int_s^{\tau_0}\!R_{\mu\nu}u^\mu u^\nu\,\dd t&=\frac{A^2|l|}{8c}\left[\tau_0-\tau-\tau\ln\frac{\tau_0}{\tau}\right]\nonumber\\&\to\frac{A^2|l|\,\tau_0}{8c}.
\end{align}
Radial timelike geodesics therefore satisfy Kr\'olak's condition, while the double integral of Tipler stays finite: the Ricci contraction along them does not by itself establish a Tipler-strong singularity. The conclusion changes for non-radial geodesics. For $f_0>0$ they cannot reach the origin, because the centrifugal term $-f_0L^2/r^2$ in $\dot r^2$ is negative and diverges. For $f_0<0$ they do reach it, with $\dot r^2\simeq|f_0|L^2/r^2$, hence $r^2\simeq2\sqrt{|f_0|}\,L\,\tau$, and the first term of the contraction dominates, $R_{\mu\nu}u^\mu u^\nu\simeq A|l|L^2/(2r^4)=A|l|/(8|f_0|\tau^2)$. Then
\begin{equation}
\int_\tau^{\tau_0}\!\dd s\int_s^{\tau_0}\!\frac{A|l|}{8|f_0|\,t^2}\,\dd t\simeq\frac{A|l|}{8|f_0|}\ln\frac{\tau_0}{\tau}\to\infty,
\end{equation}
so in the spacelike branch the double-integral criterion of Tipler is satisfied along non-radial timelike geodesics. In the marginal case $f_0=0$, $f\simeq b\,r$ with $b=A^2|l|/8$. Radial geodesics reach the center with $\dot r^2\to E^2$, so the radial result above holds with $c=E$; non-radial geodesics have $\dot r^2\simeq E^2-bL^2/r$, which becomes negative near the origin, and they do not reach it. The conclusions are therefore the following. For $m_0=0$ and every sign of $f_0$, the singularity is strong in the sense of Kr\'olak along the radial timelike geodesics that reach it. For $f_0<0$ ($AM>1$) it is, in addition, strong in the sense of Tipler along the non-radial timelike geodesics, which all reach the center. For $f_0\ge0$ the only timelike geodesics reaching the center are radial, and the Ricci contraction along them yields Kr\'olak strength. Radial null geodesics do not probe this divergence, since $R_{\mu\nu}k^\mu k^\nu=0$ by the Kerr--Schild condition.

\paragraph{Local integrability of curvature invariants.}
A complementary characterization of the central singularity
is provided by the local integrability of curvature invariants.
Unlike the Tipler and Kr\'olak criteria discussed above, which
involve curvature contractions integrated along timelike
geodesics, the present analysis concerns scalar invariants
integrated with the invariant spacetime volume measure.
Related aspects of integrable singularities have been discussed
in Refs.~\cite{Lukash:2013ts,Ovalle:2023vvu,
Antonelli:2025zxh,Estrada:2026kea}.

For the metric~\eqref{metric}, the invariant volume element is
$\sqrt{-g}\,\dd^4x=r^2\sin\theta\,
\dd t\,\dd r\,\dd\theta\,\dd\phi$.
Local integrability near the origin is therefore determined
by the radial measure $r^2\dd r$.
From Eq.~\eqref{Ricci}, the Ricci scalar behaves as
$R\sim A|l|/r^2$ and is locally integrable independently
of the central mass $m_0=M-|l|/2$.

The Kretschmann scalar $K$, being quadratic in the Riemann
tensor, provides through $\sqrt{K}$ a curvature magnitude
with the same physical dimension as $R$.
For $m_0\neq0$, its leading behavior is
$K\sim48m_0^2/r^6$, so that
$r^2\sqrt{K}\sim r^{-1}$ and the corresponding integral
diverges logarithmically.
This stronger divergence originates from the Weyl curvature.
Indeed, Eq.~\eqref{psi2} gives
$\Psi_2\sim-m_0/r^3$, whereas the Ricci curvature
remains of order $r^{-2}$.

When the central mass vanishes,
$m_0=0$, or equivalently $AM=\mathcal K$,
the leading Schwarzschild-type contribution disappears.
Using the central expansion~\eqref{f0}, one obtains
\begin{equation}
R\sim\frac{2\mathcal K}{r^2},
\qquad
K\sim\frac{4\mathcal K^2}{r^4}.
\end{equation}
The Weyl scalar is also less singular:
$\Psi_2\sim-m_1/(3r^2)$, where
$m_1=-Al/4$.
Consequently,
\begin{equation}
\int_0^\epsilon r^2|R|\,\dd r<\infty,
\qquad
\int_0^\epsilon r^2\sqrt{K}\,\dd r<\infty,
\end{equation}
for sufficiently small finite $\epsilon>0$,
while the integral of $K$ itself remains divergent.

Thus, the cancellation of the central mass removes
the nonintegrable leading contribution of the Weyl curvature,
leaving locally integrable curvature magnitudes with respect
to the spacetime volume measure.
This result holds independently of whether
$\mathcal K$ is smaller or larger than unity.
The curvature invariants nevertheless remain divergent
at the origin, and their local integrability does not
contradict the geodesic strength criteria established above,
which probe different aspects of the singularity.

%=====================================================================
\section{Thermodynamics and the inner horizon}
\label{sec:thermo}

\subsection{Surface gravity in closed form}

At any horizon $r_h$ one has $l\,\e^{-Ar_h/2}=r_h-2M$, which eliminates the exponential from $f'(r_h)$ and gives
\begin{equation}
\kappa(r_h)=\frac12 f'(r_h)=\frac{A}{4}+\frac{1-AM}{2r_h}.
\label{kappa}
\end{equation}
This expression holds for both horizons, with $\kappa_+>0$ and $\kappa_-<0$. The extremality condition $\kappa=0$ gives $r_h=2M-2/A=r_\ast$, in agreement with Proposition~1. The Hawking temperature of the event horizon is $T=\kappa_+/2\pi$, and in terms of the horizon radius and the hair parameters, using $M(r_+)=r_+/2-(l/2)\e^{-Ar_+/2}$,
\begin{equation}
T=\frac{1}{4\pi r_+}\left(1+\frac{lA}{2}\,\e^{-Ar_+/2}\right).
\label{temp}
\end{equation}
Numerical values for representative parameters are collected in Table~\ref{tab:num}.

\begin{table}[t]
\centering\small\setlength{\tabcolsep}{3.5pt}
\caption{Horizons and surface gravities ($M=1$) for representative values of the range and amplitude of the hair in the two-horizon regime. The last column gives the minimal mass $M_{\rm ext}$ of Eq.~\eqref{ext} for the same hair.}
\label{tab:num}
\begin{tabular}{ccccccc}
\toprule
$A$ & $l$ & $r_-$ & $r_+$ & $\kappa_-$ & $\kappa_+$ & $M_{\rm ext}$\\
\midrule
2 & $-2.4$ & 0.4147 & 1.4197 & $-0.7057$ & 0.1478 & 0.9377\\
3 & $-3.0$ & 0.4329 & 1.7977 & $-1.5598$ & 0.1937 & 0.8347\\
3 & $-4.0$ & 0.8060 & 1.6764 & $-0.4907$ & 0.1535 & 0.9306\\
4 & $-5.0$ & 0.6574 & 1.8847 & $-1.2816$ & 0.2041 & 0.8256\\
4 & $-8.0$ & 1.0831 & 1.7661 & $-0.3849$ & 0.1506 & 0.9431\\
\bottomrule
\end{tabular}
\end{table}

\subsection{Extended first law and Smarr relation}

The parameters $l$ and $A$ label the members of the family of solutions. In what follows they are promoted to thermodynamic variables, with conjugates defined by the derivatives of the mass function; the resulting relations are algebraic identities of the family $M(S,l,A)$, and we do not present them as a covariant derivation of the first law for the matter sector.

Using $S=\pi r_+^2$ and Eq.~\eqref{sol} at the horizon, the mass is a function $M(S,l,A)$ of the entropy and of the two hair parameters,
\begin{equation}
M=\frac{r_+}{2}-\frac l2\,\e^{-Ar_+/2},\qquad r_+=\sqrt{S/\pi}.
\end{equation}
Its differential is the extended first law,
\begin{align}
\dd M=T\,\dd S+\Phi\,\dd l+\Psi\,\dd A,\label{first}\\
\Phi=-\frac12\,\e^{-Ar_+/2},\quad \Psi=\frac{l\,r_+}{4}\,\e^{-Ar_+/2},
\end{align}
where $T=\partial M/\partial S$ coincides with $\kappa_+/2\pi$ of Eq.~\eqref{temp}, so the area law is consistent with the surface gravity of the Kerr--Schild geometry. Since $M$, $S$, $l$ and $A$ scale as $L$, $L^2$, $L$ and $L^{-1}$, Euler's theorem gives the Smarr relation
\begin{equation}
M=2TS+\Phi\,l-\Psi\,A,
\end{equation}
which holds identically. The derivatives are $T=(\partial M/\partial S)_{l,A}$, $\Phi=(\partial M/\partial l)_{S,A}$ and $\Psi=(\partial M/\partial A)_{S,l}$, and each term of the Smarr relation has dimensions of length, as $M$.

The space of solutions is three-dimensional and can be coordinatized by $(M,\sigma,A)$ or, equivalently, by $(M,l,A)$, since $l=8\pi\sigma/A$ is an invertible change of variables for $A>0$. The two hair variables can therefore be varied independently; in the variables $(l,A)$ adopted here, a change of $A$ at fixed $l$ changes the constitutive constant $\sigma$ in proportion, $\delta\sigma/\sigma=\delta A/A$, whereas a change of $l$ at fixed $A$ is a change of $\sigma$ alone. The conjugates have a transparent meaning. Since $M-m(r_+)=-(l/2)\,\e^{-Ar_+/2}$ is the mass of the fluid outside the event horizon, one has
\begin{equation}
\Phi\,l=M-m(r_+),
\end{equation}
so $\Phi$ is the exterior hair mass per unit amplitude. Similarly, $\Psi=\partial_A[M-m(r_+)]$ at fixed $r_+$ and $l$, and $\Psi\,\delta A$ is the change of the exterior hair mass produced by a change of the range, which redistributes the fluid across the horizon.

\subsection{Heat capacity and the extremal configuration}

\begin{figure*}[t]
\centering
\includegraphics[width=\textwidth]{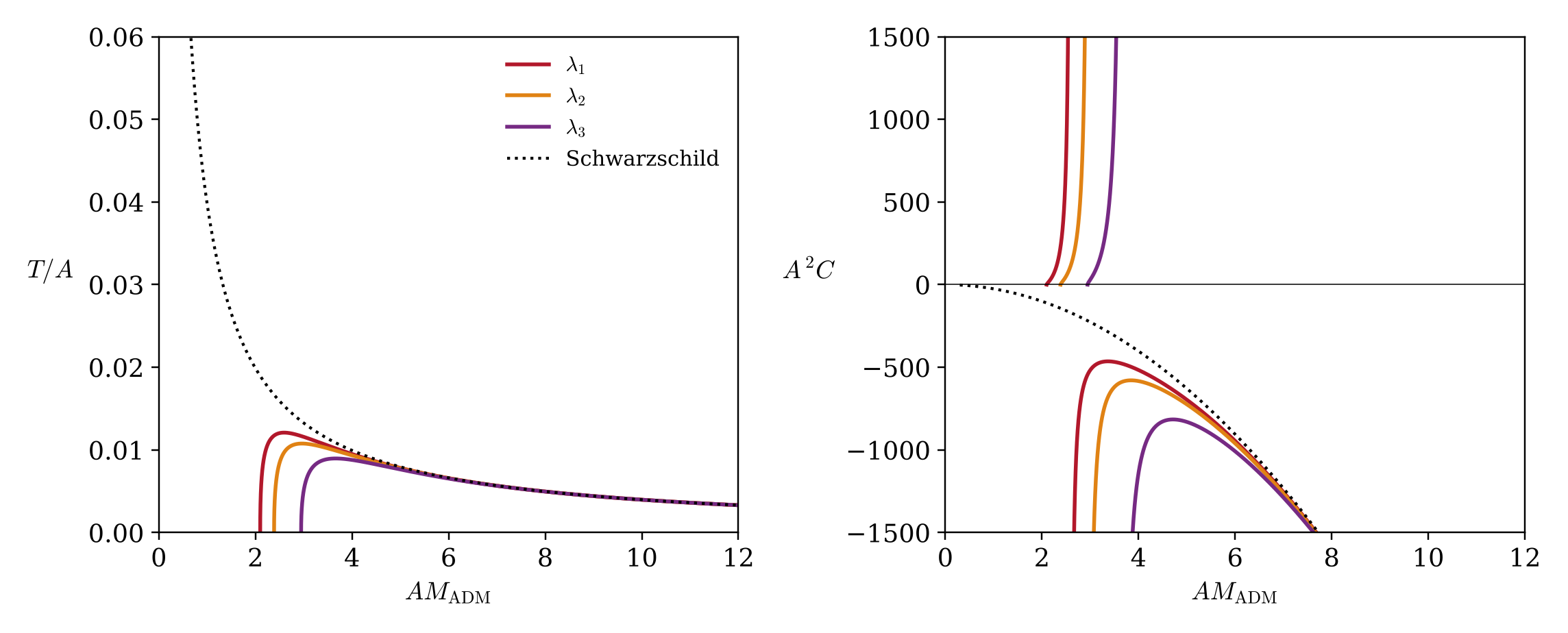}
\caption{Hawking temperature (left) and heat capacity (right) as functions of the mass for three values of the product $\lambda=lA$ with $|\lambda|>2$ (equivalently $\mathcal{K}=|\lambda|/2>1$; each curve is labeled by its value of $\lambda$)} (units $A=1$), compared with Schwarzschild (dotted). The temperature vanishes at the extremal mass, reaches a maximum and approaches the Schwarzschild curve at large mass; the heat capacity is positive between the extremal configuration and the maximum.
\label{fig:thermo}
\end{figure*}

The heat capacity at fixed hair is
\begin{equation}
C_{l,A}=T\left(\frac{\partial S}{\partial T}\right)_{l,A}=\frac{2\pi r_+\,T}{\dd T/\dd r_+}.
\label{heatcap}
\end{equation}
With $x=Ar_+/2$, Eq.~\eqref{temp} reads $T=A(1-\mathcal{K}\e^{-x})/(8\pi x)$, and
\begin{equation}
\frac{\dd T}{\dd x}=\frac{A}{8\pi x^2}\Bigl[\mathcal{K}\,\e^{-x}(1+x)-1\Bigr].
\end{equation}
Since $(1+x)\e^{-x}<1$ for $x>0$, the bracket is negative for every $x>0$ when $\mathcal{K}\le1$. For $\mathcal{K}\le1$ the temperature decreases monotonically with the mass, $C<0$, and the thermodynamic behavior is Schwarzschild-like until the central mass vanishes. For $\mathcal{K}>1$ the temperature vanishes at $M_{\rm ext}$, grows to a maximum and then decreases (Fig.~\ref{fig:thermo}). The maximum is located by $\dd T/\dd r_+=0$, which reads
\begin{equation}
1-\mathcal{K}\,\e^{-x}(1+x)=0,\qquad x=\frac{Ar_+}{2},
\end{equation}
and has the unique solution
\begin{equation}
x_c=-1-W_{-1}\!\left(-\frac{1}{\e\,\mathcal{K}}\right),
\label{lambert}
\end{equation}
where $W_{-1}$ is the lower real branch of the Lambert function \cite{Corless:1996}. Writing the condition as $u\,\e^{u}=-1/(\e\mathcal{K})$ with $u=-(1+x)$, real solutions exist for $\mathcal{K}\ge1$; for $\mathcal{K}>1$ the principal branch $W_0$ gives $u>-1$, that is $x<0$, outside the physical domain, while $W_{-1}$ gives $u<-1$ and the unique positive root, which satisfies $x_c>\ln\mathcal{K}=Ar_{\rm ext}/2$. At $\mathcal{K}=1$ the two branches meet at $u=-1$, the maximum moves to $x_c=0$, and $T\to A/(8\pi)$ there; for $\mathcal{K}<1$ there is no stationary point. At this point
\begin{equation}
M_c=\frac1A\left(x_c+\frac{1}{1+x_c}\right),\qquad T_{\max}=\frac{A}{8\pi(1+x_c)}.
\end{equation}
The heat capacity is positive for $r_{\rm ext}<r_+<2x_c/A$, diverges at $x_c$ and is negative beyond, which is a Davies-type change of local thermodynamic stability at fixed hair \cite{Davies:1977bgr}; we do not interpret the divergence as a global phase transition. Along a quasi-static sequence of configurations with $\mathcal{K}>1$ in which the mass decreases and the hair parameters are held fixed, the temperature first grows, as in Schwarzschild, until $M_c$ is reached, and then decreases to zero at $M_{\rm ext}$. The static family therefore admits a zero-temperature extremal configuration, which is a candidate remnant in an evaporation scenario that preserves the hair parameters; the actual radiative evolution also depends on the emission rates and on the response of the matter sector, which this static analysis does not specify.

\subsection{Thermodynamic topology}
\label{subsec:topological_thermodynamics}

The thermodynamic branches identified above can be further characterized by their winding numbers. We adopt the thermodynamic-defect formalism of Ref.~\cite{Wei:2022dzw}, in which equilibrium black holes correspond to zeros of a two-component vector field. This construction relates the number and character of the equilibrium states to the behavior of the Hawking temperature, including the meeting of the two branches at its maximum.

At fixed hair parameters $l$ and $A$, we introduce the off-shell free energy
\begin{equation}
 \mathcal{F}(r_+,\tau)=M(r_+)-\frac{S(r_+)}{\tau},
 \qquad S(r_+)=\pi r_+^2,
 \label{eq:top_free_energy}
\end{equation}
where $\tau>0$ is an auxiliary inverse temperature. For the positive-density branch ($l<0$), the horizon condition gives $M(r_+)=r_+/2+|l|\e^{-Ar_+/2}/2$. Using $\mathcal{K}=|l|A/2$, as in Sec.~\ref{sec:hor}, the vector field reads
\begin{align}
 \boldsymbol{\phi}&=(\phi^{r_+},\phi^\Theta),\nonumber\\
 \phi^{r_+}&=\frac{\partial\mathcal{F}}{\partial r_+}\nonumber\\
 &=\frac12\left(1-\mathcal{K}\e^{-Ar_+/2}\right)
   -\frac{2\pi r_+}{\tau},\nonumber\\
 \phi^\Theta&=-\cot\Theta\,\csc\Theta.
 \label{eq:top_phi}
\end{align}
Its zeros occur at $\Theta=\pi/2$ and $\phi^{r_+}=0$, or equivalently
\begin{equation}
 \tau=\frac{4\pi r_+}{1-\mathcal{K}\e^{-Ar_+/2}}
      =\frac{1}{T(r_+)},
 \label{eq:top_equilibrium}
\end{equation}
where $T$ is given in Eq.~\eqref{temp}. Thus each isolated zero represents an equilibrium black hole at temperature $1/\tau$. We count only outer horizons: $r_+>0$ for $0<\mathcal{K}<1$, and $r_+>r_{\rm ext}$ for $\mathcal{K}>1$, with $r_{\rm ext}$ given by Eq.~\eqref{ext}.

The winding number measures the rotation of $\boldsymbol{\phi}$ along a small counterclockwise contour enclosing an isolated zero. At $(r_i,\pi/2)$, the Jacobian is diagonal, since $\phi^{r_+}$ depends only on $r_+$ and $\phi^\Theta$ only on $\Theta$. Moreover, $\partial_\Theta\phi^\Theta|_{\pi/2}=1$. Using $\partial_{r_+}\mathcal{F}=S'(r_+)[T(r_+)-1/\tau]$ at equilibrium, we find
\begin{align}
 w_i&=\operatorname{sgn}\left.\left[
 \det\frac{\partial(\phi^{r_+},\phi^\Theta)}
 {\partial(r_+,\Theta)}\right]\right|_{(r_i,\pi/2)}
 \nonumber\\
 &=\operatorname{sgn}\left.\left[
 \frac{\dd T}{\dd r_+}\right]\right|_{r_i}.
 \label{eq:top_winding}
\end{align}
Since $C_{l,A}=T S'/T'$ and $T,S'>0$ for nonextremal outer horizons, $w_i=+1$ on the positive-heat-capacity branch and $w_i=-1$ on the negative-heat-capacity branch, as discussed in Sec.~V\,C. The former zero is locally source-like, whereas the latter has a saddle structure; a negative winding number does not imply convergence of the field from all directions.

\begin{figure*}[t]
 \centering
 \includegraphics[width=\textwidth]{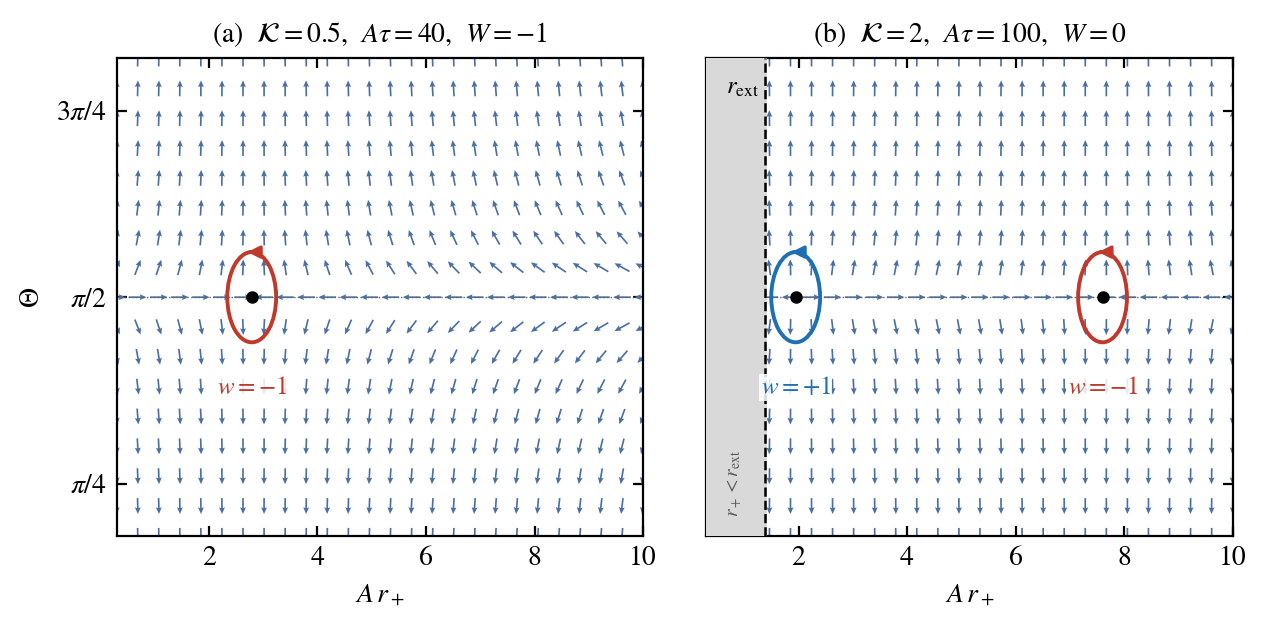}
 \caption{Normalized thermodynamic vector field $\boldsymbol{\phi}$, Eq. (\ref{eq:top_phi}), in the ($Ar_+$,$\Theta$) plane. (a) $\mathcal K=0.5$, $A\tau=40$: a single zero with $w=-1$, yielding $W=-1$. (b) $\mathcal{ K}=2$, $A\tau=100$: two zeros on the outer-horizon branch, with $w=+1$ (smaller radius) and $w=-1$ (larger radius), yielding $W=0$. The shaded region $r_+<r_{\mathrm{ext}}$ is excluded. Blue arrows indicate the direction of $\boldsymbol{\phi}$, closed loops are counterclockwise test contours, and black dots mark thermodynamic equilibrium states.}
 \label{fig:topology}
\end{figure*}

For $0<\mathcal{K}<1$, the temperature decreases monotonically from $+\infty$ as $r_+\to0^+$ to zero as $r_+\to\infty$. Accordingly, Eq.~\eqref{eq:top_equilibrium} has one positive-radius solution for every $\tau>0$. This zero has $w=-1$, giving $W=-1$. In contrast, for $\mathcal{K}>1$, the temperature vanishes at the finite extremal radius $r_{\rm ext}$, rises to $T_{\max}$, and then tends to zero at infinity. For $\tau>1/T_{\max}$, there are two equilibria: the smaller-radius solution has $w=+1$ and the larger-radius one $w=-1$, so $W=0$. The distinction between the two total winding numbers therefore follows directly from the different temperature behavior at the boundaries of the physical outer-horizon domain. Figure~\ref{fig:topology} displays the corresponding zeros of the vector field, which can be compared with the temperature and heat-capacity curves in Fig.~\ref{fig:thermo}.

When $\tau=1/T_{\max}$, the two zeros meet at the radius specified by Eq.~\eqref{lambert}. This is the Davies point, where $\dd T/\dd r_+=0$ and the heat capacity diverges. For $\tau<1/T_{\max}$, no equilibrium zeros exist in the physical domain. Thus the topological construction displays the merging and disappearance of the two equilibrium branches, without by itself establishing a global thermodynamic phase transition. The annihilation of these opposite-winding defects leaves $W=0$ unchanged; it should not be confused with the change of topological class as $\mathcal{K}$ crosses unity.

As a consistency check, the same winding signs follow from the residue prescription discussed in Ref.~\cite{Silva:2025iip}. Define
$\tau(z)=4\pi z/[1-\mathcal{K}\e^{-Az/2}]$ and $\mathcal{R}(z)=[\tau-\tau(z)]^{-1}$. At a simple positive real root $z=r_i$ of $\tau=\tau(z)$,
\begin{align}
 \operatorname{Res}_{z=r_i}\mathcal{R}(z)
 &=-\frac{1}{\tau'(r_i)}\nonumber\\
 &=\frac{1}{\tau^2 T'(r_i)}.
 \label{eq:top_residue}
\end{align}
Its sign agrees with Eq.~\eqref{eq:top_winding}. This expression applies to simple roots, not to the degenerate zero at the Davies point.

The threshold $\mathcal{K}=1$ is a boundary case. Here $T(r_+)\to A/(8\pi)$ as $r_+\to0^+$, and a negative-winding equilibrium exists only for $\tau>8\pi/A$. For $\tau<8\pi/A$, there is no equilibrium zero; at equality, the limiting zero lies on the excluded boundary $r_+=0$. Thus $W=-1$ for $0<\mathcal{K}<1$ and $W=0$ for $\mathcal{K}>1$ are the generic classifications, while $\mathcal{K}=1$ requires a boundary prescription. The change of class is tied to the appearance of a finite-radius extremal endpoint, not to the pair annihilation at the Davies point.

\subsection{The inner horizon}

\begin{figure}[t]
\centering
\includegraphics[width=\columnwidth]{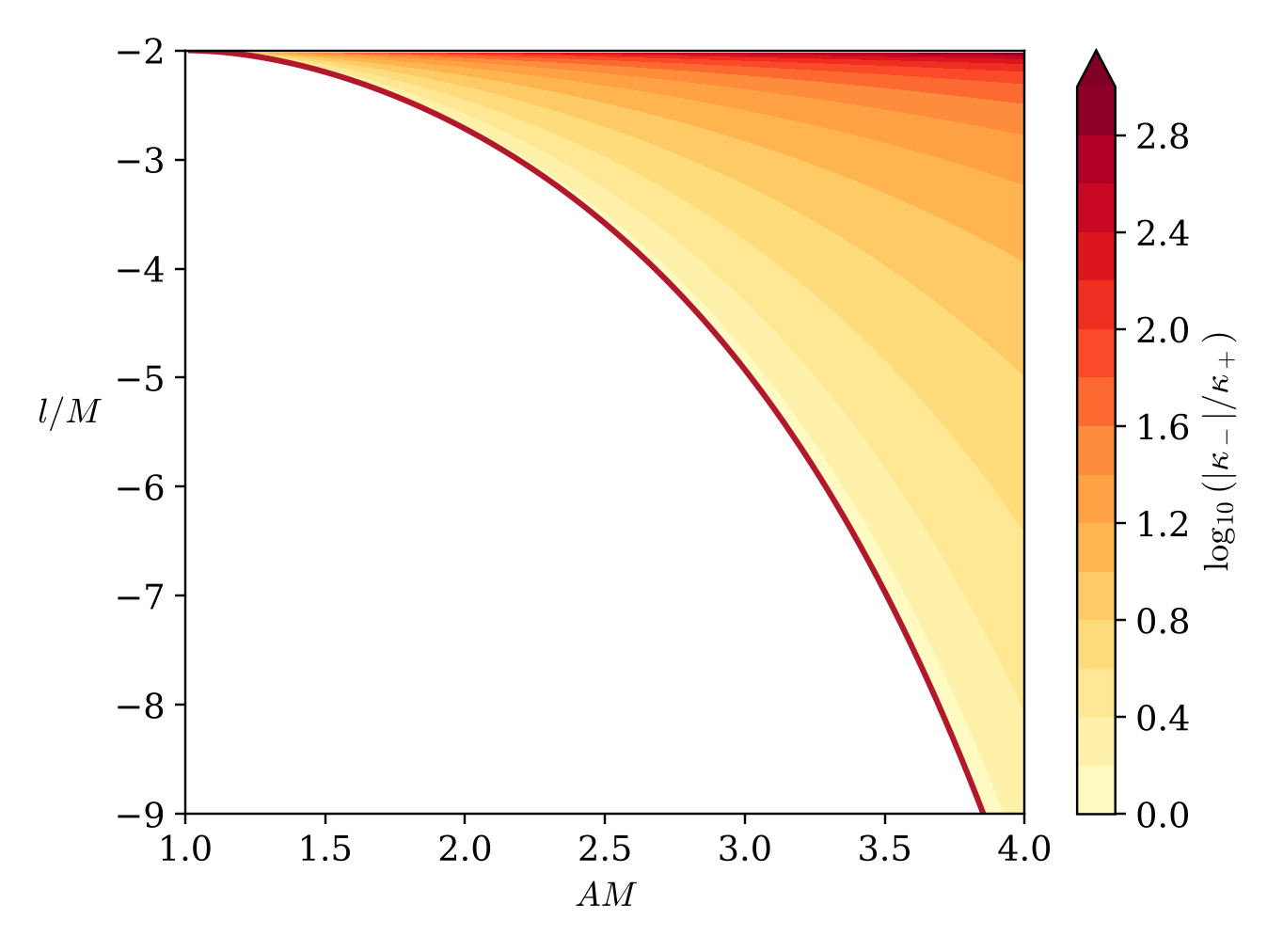}
\caption{Ratio $|\kappa_-|/\kappa_+$ in the two-horizon region of the $(AM,l/M)$ plane, on a logarithmic scale. The ratio tends to one at the extremal curve (thick) and grows without bound as the central mass approaches zero.}
\label{fig:kappa}
\end{figure}

Equation~\eqref{kappa} also controls the inner horizon. For $l_c<l<-2M$ one finds
\begin{equation}
|\kappa_-|=\frac{AM-1}{2r_-}-\frac A4,
\end{equation}
which diverges as $r_-\to0$, that is, as $l\to-2M$ from below, and tends to $\kappa_+$ at the extremal curve, where both vanish. Figure~\ref{fig:kappa} shows the ratio $|\kappa_-|/\kappa_+$, which exceeds unity throughout the two-horizon region, as in Reissner--Nordstr\"om, where it equals $(r_+/r_-)^2$. This can be proven analytically. From Eq.~\eqref{kappa},
\begin{equation}
|\kappa_-|-\kappa_+=\frac{AM-1}{2}\left(\frac{1}{r_-}+\frac{1}{r_+}-\frac{2}{r_\ast}\right),
\end{equation}
with $r_\ast=2(AM-1)/A$, so the inequality $|\kappa_-|>\kappa_+$ is equivalent to $1/r_-+1/r_+>2/r_\ast$. In terms of $z=A(r-r_\ast)/2$, the horizon condition $g(r)=l$ becomes $h(z_-)=h(z_+)$, with $h(z)=(1-z)\e^{z}$ and $z_-<0<z_+$. The function $D(z)=h(z)-h(-z)$ obeys $D(0)=0$ and $D'(z)=-z(\e^{z}-\e^{-z})<0$ for $z>0$, hence $h(z)<h(-z)$ for $z>0$, and therefore $z_+<|z_-|$, that is, $r_-+r_+<2r_\ast$. Writing $r_\pm=r_\ast+\delta_\pm$ with $\delta_-<0<\delta_+$ and $\delta_-+\delta_+<0$,
\begin{equation}
\frac{1}{r_-}+\frac{1}{r_+}-\frac{2}{r_\ast}=-\frac{r_\ast(\delta_-+\delta_+)+2\delta_-\delta_+}{r_\ast\,r_-r_+}>0,
\end{equation}
since both terms in the numerator are negative. Hence $|\kappa_-|>\kappa_+$ for every $l_c<l<-2M$. At the extremal curve $z_\pm\to0$ symmetrically and the ratio tends to one, while as $l\to-2M$ one has $r_-\to0$ and the ratio diverges; the numerical values of Table~\ref{tab:num} lie in between. The blueshift factor of infalling radiation at the Cauchy horizon grows as $\e^{|\kappa_-|v}$ in the static background. This is the kinematic ingredient of the mass-inflation mechanism \cite{Poisson:1990eh,Ori:1991zz}, and its strength can be tuned by the hair parameters; it does not by itself establish mass inflation, which depends on the dynamics of the perturbations and on the response of the matter sector.

%=====================================================================
\section{A family of nonlocal closures and a regular member}
\label{sec:family}

The closure~\eqref{nleos} uses the volume weight $x^2$ in the integral. More generally one can weight the enclosed density with a power of the radius,
\begin{equation}
p_r=\rho+\frac{A}{r^k}\int_0^r x^k\rho(x)\,\dd x+\frac{\sigma}{r^k},\qquad k<3,
\label{nleosk}
\end{equation}
and the case $k=2$ is the one of the previous sections. The other members form a postulated family of closures, not a prediction of the case $k=2$. The bound $k<3$ is the condition for a finite central mass: the density behaves as $r^{-k}$ near the origin, so $m'=4\pi r^2\rho\propto r^{2-k}$ is integrable at $r=0$ if and only if $k<3$, which is also the condition for $\Gamma(3-k)$ to be finite. Positivity of the density requires $\sigma<0$ for every $k$, and regularity at the center requires in addition $m(0)=0$ and $k\le0$, as shown below. The Kerr--Schild condition $p_r=-\rho$ gives for the enclosed quantity $\mathcal{Q}_k=\int_0^r x^k\rho\,\dd x$ the same linear equation~\eqref{Qode}, so that $r^k\rho=\mathcal{Q}_k'=-(\sigma/2)\,\e^{-Ar/2}$. The density, the mass function and the metric function are
\begin{align}
\rho&=-\frac{\sigma}{2}\,\frac{\e^{-Ar/2}}{r^k},\\
p_t&=\left(\frac k2-1+\frac{Ar}{4}\right)\rho,\\
m(r)&=M+\frac{l_k}{2}\,\Gamma\!\left(3-k,\tfrac{Ar}{2}\right),\\
f(r)&=1-\frac{2M}{r}-\frac{l_k}{r}\,\Gamma\!\left(3-k,\tfrac{Ar}{2}\right),
\label{fk}
\end{align}
with $l_k=4\pi\sigma\,(2/A)^{3-k}$ and $\Gamma(a,z)$ the upper incomplete gamma function. The case $k=2$ reproduces Eq.~\eqref{sol}, since $\Gamma(1,z)=\e^{-z}$. For $k=1$ and $k=0$ one has $\Gamma(2,z)=(1+z)\e^{-z}$ and $\Gamma(3,z)=(2+2z+z^2)\e^{-z}$, so the profiles are Yukawa terms with polynomial prefactors. The asymptotic mass is $M$ and the hair is screened exponentially for every $k$.
\begin{figure*}
\centering
\includegraphics[width=\textwidth]{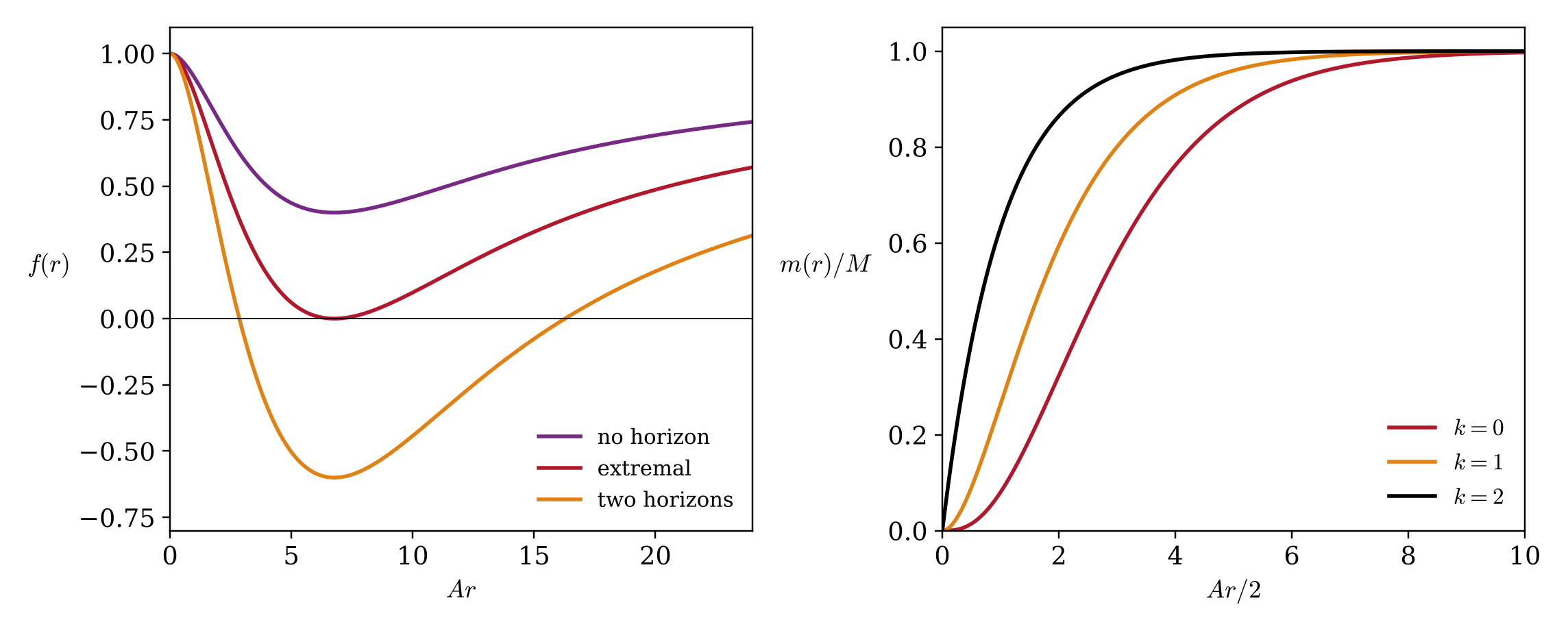}
\caption{Left: metric function of the regular member $k=0$ for masses below, at and above the extremal value; the de~Sitter core is visible at $r\to0$. Right: mass function normalized to $M$ for $k=0,1,2$ with $m(0)=0$; its growth near the origin is $r^{3-k}$.}
\label{fig:family}
\end{figure*}

The weight controls the central region. Imposing $m(0)=0$, which fixes $l_k=-2M/\Gamma(3-k)$, the mass function is the regularized lower incomplete gamma function,
\begin{equation}
m(r)=M\,P\!\left(3-k,\tfrac{Ar}{2}\right)\simeq M\,\frac{(Ar/2)^{3-k}}{\Gamma(4-k)},
\end{equation}
that is, $m\propto r^{3-k}$ near the origin (Fig.~\ref{fig:family}, right). For $k=2$ this is the central region of Sect.~\ref{sec:core}, a global monopole for $AM<1$, for $k=1$ the Ricci scalar diverges as $1/r$ and the Kretschmann scalar as $1/r^2$, and for $k=0$ the mass function grows as $r^3$ and the density is finite at the center, $\rho(0)=MA^3/(64\pi)$ with $p_r=p_t=-\rho$ there. The metric function behaves as
\begin{equation}
f(r)=1-\frac{MA^3}{24}\,r^2+\mathcal{O}(r^3),
\end{equation}
which is the de~Sitter core with effective cosmological constant $\Lambda_{\rm eff}=MA^3/8=8\pi\rho(0)$, and all curvature invariants are finite: since $m=\mathcal{O}(r^3)$, the quantities $f''$, $f'/r$ and $(1-f)/r^2$ that build the Kretschmann scalar remain bounded, with $K\to8\Lambda_{\rm eff}^2/3$ and $R\to4\Lambda_{\rm eff}$ at the center. The member $k=0$ is therefore a regular black hole,
\begin{equation}
f(r)=1-\frac{2M}{r}\left[1-\left(1+\frac{Ar}{2}+\frac{A^2r^2}{8}\right)\e^{-Ar/2}\right],
\label{regular}
\end{equation}
in the family of regular black holes with a de~Sitter core \cite{Dymnikova:1992ux,AyonBeato:1998ub,Hayward:2005gi}, but obtained here from a nonlocal closure of an anisotropic fluid, without a nonlinear electrodynamics source. The strong energy condition is violated in the core, as is required by any regular core of this type.

The horizons of the regular member are determined by $AM=x/P(3,x)$ with $x=Ar/2$, and the degenerate horizon by the additional condition $(x^3+x^2+2x+2)\,\e^{-x}=2$. This has a single root, $x_{\rm ext}=3.3836$, which gives
\begin{equation}
AM_{\rm ext}=5.1494,\qquad Ar_{\rm ext}=6.7673.
\end{equation}
A horizon exists if and only if $AM\ge AM_{\rm ext}$; for larger masses there are two horizons, $r_-<r_+$, and at $AM=AM_{\rm ext}$ they merge at $r_{\rm ext}$. The metric function is positive for $r<r_-$, negative for $r_-<r<r_+$ and positive for $r>r_+$. The inner horizon $r_-$ is a Killing horizon that bounds the static de~Sitter core; seen from the core it plays the role of the cosmological horizon of de~Sitter space, with $f'(r_-)<0$, while seen from outside it is the Cauchy horizon of the black hole. The outer horizon $r_+$ is the event horizon (Fig.~\ref{fig:family}, left), with the causal structure of the Hayward black hole.

The energy conditions for the whole family follow from $\rho\propto-\sigma$, $p_r=-\rho$ and $p_t/\rho=k/2-1+Ar/4$. For $\sigma<0$ and $0\le k<3$: the weak and null energy conditions hold at every radius, since $\rho>0$ and $\rho+p_t=(k/2+Ar/4)\rho\ge0$; the strong energy condition, $\rho+p_r+2p_t=2p_t\ge0$, holds for $Ar\ge4-2k$ and is violated closer to the center unless $k\ge2$; and the dominant energy condition, $|p_t|\le\rho$, holds for $Ar\le8-2k$. For $k=2$ these reduce to the statements of Sect.~\ref{sec:sol}. For $k<0$ the null energy condition is violated near the origin.

%=====================================================================
\section{Conclusions}
\label{sec:concl}

We have shown that a single anisotropic fluid, constrained by the Kerr--Schild condition and by a nonlocal equation of state, determines a neutral black hole with a finite-range hair of Yukawa type. The profile is not chosen: it follows from the closure, which reduces the problem to a linear equation for the enclosed matter. This distinguishes the construction from earlier Yukawa-type black holes, where the profile comes from a prescribed potential or from the additional degrees of freedom of the gravitational theory, and it places the solution in the class of exact Einstein solutions with an effective source whose constitutive relation is explicit.

The nonlocal closure selects a neutral family with an analytic two-horizon region and an analytic extremal curve. The amplitude and the range of the hair generate the whole sequence of causal structures, from the Schwarzschild-like case to configurations with a Cauchy horizon, extremal black holes and naked singularities. The boundaries between these regimes, the mass and the radius of the zero-temperature extremal configuration, the surface gravities of both horizons and the point at which the heat capacity changes sign are all available in closed form, a rarity for a metric whose horizon equation is transcendental. The results split naturally into two groups. The horizon classification, the surface gravities, the temperature and the extended first law depend only on the form of the metric function, and would hold for any source producing it; the closure is what makes this metric the unique output of a constitutive relation, and it alone fixes the sign and shape of the density, the anisotropy, the energy conditions and the family of weights. The inner horizon is non-degenerate, its surface gravity provably exceeds that of the event horizon in the whole two-horizon region, and its blueshift factor can be enhanced by the hair, which makes the neutral configurations a controlled setting in which the kinematic ingredients of the mass-inflation mechanism can be varied without an electromagnetic field. In the regime in which the Schwarzschild term is cancelled, the center is a global monopole whose strength is fixed by the hair where the geometry is static there, and a homogeneous geometry of Kantowski--Sachs type where it is not; the singularity is far milder than in Schwarzschild, satisfies Kr\'olak's criterion, and is strong in the sense of Tipler along non-radial geodesics in the spacelike branch. When the central mass vanishes, the leading nonintegrable Weyl contribution is also removed, so that the square root of the Kretschmann scalar becomes, like the Ricci scalar, locally integrable with respect to the invariant spacetime volume, despite the divergence of both at the origin.

The extended thermodynamics, with the amplitude and the range of the hair as variables, is consistent with the area law and yields a Smarr relation, together with a zero-temperature extremal configuration that is a candidate remnant under an evaporation scenario that preserves the hair. The thermodynamic topological analysis further revealed two distinct classes of equilibrium configurations, governed by the strength and radial decay of the gravitational hair. In the Schwarzschild-like regime, a single defect with negative winding number characterizes the thermodynamic equilibrium, whereas the regime admitting an extremal configuration exhibits two defects with opposite winding numbers and vanishing total topological charge. Their merging at the Davies point, where the heat capacity diverges, provides a topological description of the change in local thermodynamic stability.

Perhaps the most distinctive outcome is that the closure is the first member of a family labeled by the weight of the nonlocal integral, in which the weight alone controls the central region: the same constitutive principle passes continuously from the singular, monopole-like core of the volume weight to a regular black hole with a de~Sitter core for the flat weight, obtained without nonlinear electrodynamics and with the causal structure of the Hayward black hole. In this construction the regularity of the core is a statement about the constitutive relation, not about an ad hoc choice of the mass function.

The natural next steps share a common requirement, a dynamical model of the matter sector. Such a model would allow the radial and gravitational stability of the configurations and of the Cauchy horizon to be analyzed, the radiative evolution toward the extremal configuration to be followed, and a covariant derivation of the first law for the hair variables to be given; it would also address the microscopic origin of the nonlocal relation, and a variational formulation of the matter sector, if it exists, would turn the conjugates of the hair variables into Noether charges. On the observational side, the monopole-like core and the screened hair suggest looking for signatures that discriminate between the volume-weighted and the regular members of the family. The family of closures and the Kerr--Schild form further invite the construction of rotating generalizations.

\bibliography{biblio}

@article{An:2021plu,
    author = "An, Yu-Sen and Li, Li and Yang, Fu-Guo",
    title = "{No Cauchy horizon theorem for nonlinear electrodynamics black holes with charged scalar hairs}",
    eprint = "2106.01069",
    archivePrefix = "arXiv",
    primaryClass = "gr-qc",
    doi = "10.1103/PhysRevD.104.024040",
    journal = "Phys. Rev. D",
    volume = "104",
    number = "2",
    pages = "024040",
    year = "2021"
}

@article{Antonelli:2025zxh,
    author = "Antonelli, Tommaso and Sebastianutti, Marco",
    title = "{Singularity and differentiability at the origin of static and spherically symmetric black holes}",
    eprint = "2509.15477",
    archivePrefix = "arXiv",
    primaryClass = "gr-qc",
    doi = "10.1103/hf4r-19xh",
    journal = "Phys. Rev. D",
    volume = "113",
    number = "6",
    pages = "064007",
    year = "2026"
}

@article{AraujoFilho:2023ibe,
    author = "Ara{\'u}jo Filho, A. A. and Jusufi, Kimet and Cuadros-Melgar, B. and Leon, Genly",
    title = "{Dark matter signatures of black holes with Yukawa potential}",
    journal = "Phys. Dark Univ.",
    volume = "44",
    pages = "101500",
    doi = "10.1016/j.dark.2024.101500",
    eprint = "2310.17081",
    archivePrefix = "arXiv",
    primaryClass = "gr-qc",
    year = "2023"
}

@article{AraujoFilho:2024ctw,
    author = "Ara{\'u}jo Filho, A. A. and Jusufi, Kimet and Cuadros-Melgar, B. and Leon, Genly and Jawad, Abdul and Pellicer, C. E.",
    title = "{Charged black holes with Yukawa potential}",
    journal = "Phys. Dark Univ.",
    volume = "46",
    pages = "101711",
    doi = "10.1016/j.dark.2024.101711",
    eprint = "2401.15211",
    archivePrefix = "arXiv",
    primaryClass = "gr-qc",
    year = "2024"
}

@article{AyonBeato:1998ub,
    author = "Ay{\'o}n-Beato, Eloy and Garc{\'\i}a, Alberto",
    title = "{Regular black hole in general relativity coupled to nonlinear electrodynamics}",
    journal = "Phys. Rev. Lett.",
    volume = "80",
    pages = "5056--5059",
    year = "1998"
}

@article{Bardeen:1973gs,
    author = "Bardeen, James M. and Carter, B. and Hawking, S. W.",
    title = "{The Four Laws of Black Hole Mechanics}",
    journal = "Commun. Math. Phys.",
    volume = "31",
    pages = "161--170",
    year = "1973",
    doi = "10.1007/BF01645742"
}

@article{Barriola:1989hx,
    author = "Barriola, Manuel and Vilenkin, Alexander",
    title = "{Gravitational field of a global monopole}",
    journal = "Phys. Rev. Lett.",
    volume = "63",
    pages = "341--343",
    year = "1989",
    doi = "10.1103/PhysRevLett.63.341"
}

@article{Bekenstein:1973ur,
    author = "Bekenstein, Jacob D.",
    title = "{Black Holes and Entropy}",
    journal = "Phys. Rev. D",
    volume = "7",
    pages = "2333--2346",
    year = "1973",
    doi = "10.1103/PhysRevD.7.2333"
}

@article{Bekenstein:1995un,
    author = "Bekenstein, Jacob D.",
    title = "{Novel no-scalar-hair theorem for black holes}",
    journal = "Phys. Rev. D",
    volume = "51",
    pages = "R6608--R6611",
    year = "1995",
    doi = "10.1103/PhysRevD.51.R6608"
}

@article{Bonanno:2000ep,
    author = "Bonanno, Alfio and Reuter, Martin",
    title = "{Renormalization group improved black hole space-times}",
    eprint = "hep-th/0002196",
    archivePrefix = "arXiv",
    reportNumber = "INFN-CT-03-00, MZ-TH-00-04",
    doi = "10.1103/PhysRevD.62.043008",
    journal = "Phys. Rev. D",
    volume = "62",
    pages = "043008",
    year = "2000"
}

@article{Cai:2020wrp,
    author = "Cai, Rong-Gen and Li, Li and Yang, Run-Qiu",
    title = "{No Inner-Horizon Theorem for Black Holes with Charged Scalar Hairs}",
    eprint = "2009.05520",
    archivePrefix = "arXiv",
    primaryClass = "gr-qc",
    doi = "10.1007/JHEP03(2021)263",
    journal = "JHEP",
    volume = "03",
    pages = "263",
    year = "2021"
}

@article{Carballo-Rubio:2018pmi,
    author = "Carballo-Rubio, Ra\'ul and Di Filippo, Francesco and Liberati, Stefano and Pacilio, Costantino and Visser, Matt",
    title = "{On the viability of regular black holes}",
    eprint = "1805.02675",
    archivePrefix = "arXiv",
    primaryClass = "gr-qc",
    doi = "10.1007/JHEP07(2018)023",
    journal = "JHEP",
    volume = "07",
    pages = "023",
    year = "2018"
}

@article{Carter:1971zc,
    author = "Carter, Brandon",
    title = "{Axisymmetric Black Hole Has Only Two Degrees of Freedom}",
    doi = "10.1103/PhysRevLett.26.331",
    journal = "Phys. Rev. Lett.",
    volume = "26",
    pages = "331--333",
    year = "1971"
}

@article{Chrusciel:2012jk,
    author = "Chru\'sciel, Piotr T. and Costa, Jo\~ao Louren\c{c}o and Heusler, Markus",
    title = "{Stationary Black Holes: Uniqueness and Beyond}",
    eprint = "1205.6112",
    archivePrefix = "arXiv",
    primaryClass = "gr-qc",
    doi = "10.12942/lrr-2012-7",
    journal = "Living Rev. Rel.",
    volume = "15",
    pages = "7",
    year = "2012"
}

@article{Clarke:1985jg,
    author = "Clarke, C. J. S. and Kr{\'o}lak, A.",
    title = "{Conditions for the occurrence of strong curvature singularities}",
    journal = "J. Geom. Phys.",
    volume = "2",
    pages = "127--143",
    year = "1985"
}

@article{Corless:1996,
    author = "Corless, R. M. and Gonnet, G. H. and Hare, D. E. G. and Jeffrey, D. J. and Knuth, D. E.",
    title = "{On the Lambert $W$ function}",
    journal = "Adv. Comput. Math.",
    volume = "5",
    pages = "329--359",
    year = "1996"
}

@article{Curiel:2014zba,
    author = "Curiel, Erik",
    title = "{A Primer on Energy Conditions}",
    archivePrefix = "arXiv",
    primaryClass = "physics.hist-ph",
    doi = "10.1007/978-1-4939-3210-8_3",
    journal = "Einstein Stud.",
    volume = "13",
    pages = "43--104",
    year = "2017"
}

@article{Dafermos:2003vim,
    author = "Dafermos, Mihalis",
    title = "{Stability and Instability of the Cauchy Horizon for the Spherically Symmetric Einstein-Maxwell-Scalar Field Equations}",
    journal = "Ann. Math",
    volume = "158",
    number = "3",
    pages = "875--928",
    year = "2003"
}

@article{Davies:1977bgr,
    author = "Davies, P. C. W.",
    title = "{Thermodynamic Phase Transitions of Kerr-Newman Black Holes in de Sitter Space}",
    journal = "Proc. Roy. Soc. Lond. A",
    volume = "353",
    pages = "499--521",
    year = "1977",
    doi = "10.1098/rspa.1977.0047"
}

@article{Devecioglu:2021xug,
    author = "Devecioglu, Deniz O. and Park, Mu-In",
    title = "{No scalar-haired Cauchy horizon theorem in Einstein-Maxwell-Horndeski theories}",
    eprint = "2101.10116",
    archivePrefix = "arXiv",
    primaryClass = "hep-th",
    doi = "10.1016/j.physletb.2022.137107",
    journal = "Phys. Lett. B",
    volume = "829",
    pages = "137107",
    year = "2022"
}

@article{Dymnikova:1992ux,
    author = "Dymnikova, Irina",
    title = "{Vacuum nonsingular black hole}",
    journal = "Gen. Rel. Grav.",
    volume = "24",
    pages = "235--242",
    year = "1992"
}

@article{Estrada:2026kea,
    author = "Estrada, Milko",
    title = {{Reissner Nordstr{\"o}m black holes with integrable singularity interiors supported by string distributions}},
    eprint = "2602.03050",
    archivePrefix = "arXiv",
    primaryClass = "gr-qc",
    doi = "10.1140/epjc/s10052-026-15888-6",
    journal = "Eur. Phys. J. C",
    volume = "86",
    number = "7",
    pages = "774",
    year = "2026"
}

@article{Fielder:2020lpk,
    author = "Fielder, Catherine E. and Mao, Yao-Yuan and Zentner, Andrew R. and Newman, Jeffrey A. and Wu, Hao-Yi and Wechsler, Risa",
    title = "{Illuminating Dark Matter Halo Density Profiles Without Subhaloes}",
    eprint = "2007.02964",
    archivePrefix = "arXiv",
    primaryClass = "astro-ph.GA",
    doi = "10.1093/mnras/staa2851",
    journal = "Mon. Not. Roy. Astron. Soc.",
    volume = "499",
    number = "2",
    pages = "2426--2444",
    year = "2020"
}

@article{German:2001ti,
    author = "German, Wenceslao S. and Moss, Ian G.",
    title = "{Cauchy horizon stability and cosmic censorship}",
    eprint = "gr-qc/0103080",
    archivePrefix = "arXiv",
    doi = "10.1088/0264-9381/18/23/306",
    journal = "Class. Quant. Grav.",
    volume = "18",
    pages = "5097--5102",
    year = "2001"
}

@article{Gonzalez:2023rsd,
    author = "Gonzalez, Esteban and Jusufi, Kimet and Leon, Genly and Saridakis, Emmanuel N.",
    title = "{Observational constraints on Yukawa cosmology and connection with black hole shadows}",
    journal = "Phys. Dark Univ.",
    volume = "42",
    pages = "101304",
    doi = "10.1016/j.dark.2023.101304",
    eprint = "2305.14305",
    archivePrefix = "arXiv",
    primaryClass = "gr-qc",
    year = "2023"
}

@article{Gurses:1975vu,
    author = "G{\"u}rses, M. and G{\"u}rsey, F.",
    title = "{Lorentz covariant treatment of the Kerr--Schild geometry}",
    journal = "J. Math. Phys.",
    volume = "16",
    pages = "2385--2390",
    year = "1975"
}

@article{Hajian:2023cwe,
    author = "Hajian, Kamal and Tekin, Bayram",
    title = "{Coupling Constants as Conserved Charges in Black Hole Thermodynamics}",
    journal = "Phys. Rev. Lett.",
    volume = "132",
    number = "19",
    pages = "191401",
    year = "2024",
    doi = "10.1103/PhysRevLett.132.191401"
}

@article{Hartnoll:2020rwq,
    author = "Hartnoll, Sean A. and Horowitz, Gary T. and Kruthoff, Jorrit and Santos, Jorge E.",
    title = "{Gravitational duals to the grand canonical ensemble abhor Cauchy horizons}",
    eprint = "2006.10056",
    archivePrefix = "arXiv",
    primaryClass = "hep-th",
    doi = "10.1007/JHEP10(2020)102",
    journal = "JHEP",
    volume = "10",
    pages = "102",
    year = "2020"
}

@article{Hawking:1971vc,
    author = "Hawking, S. W.",
    title = "{Black holes in general relativity}",
    doi = "10.1007/BF01877517",
    journal = "Commun. Math. Phys.",
    volume = "25",
    pages = "152--166",
    year = "1972"
}

@book{Hawking:1973uf,
    author = "Hawking, S. W. and Ellis, G. F. R.",
    title = "{The Large Scale Structure of Space-Time}",
    doi = "10.1017/CBO9780511524646",
    isbn = "978-0-521-20016-5",
    publisher = "Cambridge University Press",
    series = "Cambridge Monographs on Mathematical Physics",
    month = "2",
    year = "1973"
}

@article{Hawking:1975vcx,
    author = "Hawking, S. W.",
    title = "{Particle creation by black holes}",
    journal = "Commun. Math. Phys.",
    volume = "43",
    pages = "199--220",
    year = "1975"
}

@article{Hayward:2005gi,
    author = "Hayward, Sean A.",
    title = "{Formation and evaporation of regular black holes}",
    journal = "Phys. Rev. Lett.",
    volume = "96",
    pages = "031103",
    year = "2006",
    eprint = "gr-qc/0506126",
    archivePrefix = "arXiv"
}

@article{Herdeiro:2015waa,
    author = "Herdeiro, Carlos A. R. and Radu, Eugen",
    title = "{Asymptotically flat black holes with scalar hair: a review}",
    eprint = "1504.08209",
    archivePrefix = "arXiv",
    primaryClass = "gr-qc",
    doi = "10.1142/S0218271815420146",
    journal = "Int. J. Mod. Phys. D",
    volume = "24",
    number = "09",
    pages = "1542014",
    year = "2015"
}

@article{Hernandez:1998hk,
    author = "Hernandez, H. and Nunez, L. A. and Percoco, U.",
    title = "{Nonlocal equation of state in general relativistic radiating spheres}",
    eprint = "gr-qc/9806029",
    archivePrefix = "arXiv",
    journal = "Class. Quant. Grav.",
    volume = "16",
    pages = "871--896",
    year = "1999",
    doi = "10.1088/0264-9381/16/3/017"
}

@article{Hernandez:2001nr,
    author = "Hernandez, H. and Nu\~{n}ez, L. A.",
    title = "{Nonlocal equation of state in anisotropic static fluid spheres in general relativity}",
    eprint = "gr-qc/0107025",
    archivePrefix = "arXiv",
    doi = "10.1139/p03-124",
    journal = "Can. J. Phys.",
    volume = "82",
    pages = "29--51",
    year = "2004"
}

@article{Heydarzade:2023dof,
    author = "Heydarzade, Yaghoub and Misyura, Maxim and Vertogradov, Vitalii",
    title = "{Hairy Kiselev black hole solutions}",
    eprint = "2307.04556",
    archivePrefix = "arXiv",
    primaryClass = "gr-qc",
    doi = "10.1103/PhysRevD.108.044073",
    journal = "Phys. Rev. D",
    volume = "108",
    number = "4",
    pages = "044073",
    year = "2023"
}

@article{Israel:1967wq,
    author = "Israel, Werner",
    title = "{Event horizons in static vacuum space-times}",
    doi = "10.1103/PhysRev.164.1776",
    journal = "Phys. Rev.",
    volume = "164",
    pages = "1776--1779",
    year = "1967"
}

@article{Iyer:1994ys,
    author = "Iyer, Vivek and Wald, Robert M.",
    title = "{Some Properties of Noether Charge and a Proposal for Dynamical Black Hole Entropy}",
    eprint = "gr-qc/9403028",
    archivePrefix = "arXiv",
    journal = "Phys. Rev. D",
    volume = "50",
    pages = "846--864",
    year = "1994",
    doi = "10.1103/PhysRevD.50.846"
}

@article{Jacobson:2007tj,
    author = "Jacobson, Ted",
    title = "{When is g(tt) g(rr) = -1?}",
    eprint = "0707.3222",
    archivePrefix = "arXiv",
    primaryClass = "gr-qc",
    doi = "10.1088/0264-9381/24/22/N02",
    journal = "Class. Quant. Grav.",
    volume = "24",
    pages = "5717--5719",
    year = "2007"
}

@article{Kerr:1963ud,
    author = "Kerr, Roy P.",
    title = "{Gravitational field of a spinning mass as an example of algebraically special metrics}",
    doi = "10.1103/PhysRevLett.11.237",
    journal = "Phys. Rev. Lett.",
    volume = "11",
    pages = "237--238",
    year = "1963"
}

@article{Kerr:1965wfc,
    author = "Kerr, Roy Patrick and Schild, Alfred",
    title = "{Some algebraically degenerate solutions of Einstein\textquoteright{}s gravitational field equations}",
    journal = "Proc. Symp. Appl. Math.",
    volume = "17",
    pages = "199",
    year = "1965"
}

@article{Kiselev:2002dx,
    author = "Kiselev, V. V.",
    title = "{Quintessence and black holes}",
    eprint = "gr-qc/0210040",
    archivePrefix = "arXiv",
    doi = "10.1088/0264-9381/20/6/310",
    journal = "Class. Quant. Grav.",
    volume = "20",
    pages = "1187--1198",
    year = "2003"
}

@article{Konoplya:2019sns,
    author = "Konoplya, R. A.",
    title = "{Shadow of a black hole surrounded by dark matter}",
    eprint = "1905.00064",
    archivePrefix = "arXiv",
    primaryClass = "gr-qc",
    doi = "10.1016/j.physletb.2019.05.043",
    journal = "Phys. Lett. B",
    volume = "795",
    pages = "1--6",
    year = "2019"
}

@article{Konoplya:2022hbl,
    author = "Konoplya, R. A. and Zhidenko, A.",
    title = "{Solutions of the Einstein Equations for a Black Hole Surrounded by a Galactic Halo}",
    eprint = "2202.02205",
    archivePrefix = "arXiv",
    primaryClass = "gr-qc",
    doi = "10.3847/1538-4357/ac76bc",
    journal = "Astrophys. J.",
    volume = "933",
    pages = "166",
    year = "2022"
}

@article{Krolak:1987,
    author = "Kr{\'o}lak, A.",
    title = "{Strong curvature singularities and causal simplicity}",
    journal = "J. Math. Phys.",
    volume = "28",
    pages = "138--141",
    year = "1987"
}

@misc{Li:2026drw,
    author = "Li, Haida and Zhang, Xiangdong",
    title = "{Extended spherically symmetric solutions in revised Deser--Woodard nonlocal gravity}",
    eprint = "2606.22086",
    archivePrefix = "arXiv",
    primaryClass = "gr-qc",
    year = "2026"
}

@article{Lu:2015cqa,
    author = "L{\"u}, H. and Perkins, A. and Pope, C. N. and Stelle, K. S.",
    title = "{Black Holes in Higher-Derivative Gravity}",
    journal = "Phys. Rev. Lett.",
    volume = "114",
    pages = "171601",
    year = "2015",
    eprint = "1502.01028",
    archivePrefix = "arXiv",
    primaryClass = "hep-th"
}

@article{Lu:2015psa,
    author = "L{\"u}, H. and Perkins, A. and Pope, C. N. and Stelle, K. S.",
    title = "{Spherically Symmetric Solutions in Higher-Derivative Gravity}",
    journal = "Phys. Rev. D",
    volume = "92",
    pages = "124019",
    doi = "10.1103/PhysRevD.92.124019",
    year = "2015",
    eprint = "1508.00010",
    archivePrefix = "arXiv",
    primaryClass = "gr-qc"
}

@article{Lukash:2013ts,
    author = "Lukash, Vladimir N. and Strokov, Vladimir N.",
    title = "{Space-Times with Integrable Singularity}",
    eprint = "1301.5544",
    archivePrefix = "arXiv",
    primaryClass = "gr-qc",
    doi = "10.1142/S0217751X13500073",
    journal = "Int. J. Mod. Phys. A",
    volume = "28",
    pages = "1350007",
    year = "2013"
}

@article{Martinez:2002ru,
    author = "Martinez, Cristian and Troncoso, Ricardo and Zanelli, Jorge",
    title = "{De Sitter black hole with a conformally coupled scalar field in four-dimensions}",
    eprint = "hep-th/0205319",
    archivePrefix = "arXiv",
    reportNumber = "CECS-PHY-02-05",
    doi = "10.1103/PhysRevD.67.024008",
    journal = "Phys. Rev. D",
    volume = "67",
    pages = "024008",
    year = "2003"
}

@article{Martinez:2004nb,
    author = "Martinez, Cristian and Troncoso, Ricardo and Zanelli, Jorge",
    title = "{Exact black hole solution with a minimally coupled scalar field}",
    eprint = "hep-th/0406111",
    archivePrefix = "arXiv",
    reportNumber = "CECS-PHY-04-12",
    doi = "10.1103/PhysRevD.70.084035",
    journal = "Phys. Rev. D",
    volume = "70",
    pages = "084035",
    year = "2004"
}

@article{Mayo:1996mv,
    author = "Mayo, Avraham E. and Bekenstein, Jacob D.",
    title = "{No hair for spherical black holes: Charged and nonminimally coupled scalar field with self-interaction}",
    journal = "Phys. Rev. D",
    volume = "54",
    pages = "5059--5069",
    year = "1996",
    doi = "10.1103/PhysRevD.54.5059"
}

@article{Newman:1965my,
    author = "Newman, E T. and Couch, R. and Chinnapared, K. and Exton, A. and Prakash, A. and Torrence, R.",
    title = "{Metric of a Rotating, Charged Mass}",
    doi = "10.1063/1.1704351",
    journal = "J. Math. Phys.",
    volume = "6",
    pages = "918--919",
    year = "1965"
}

@article{Ori:1991zz,
    author = "Ori, Amos",
    title = "{Inner structure of a charged black hole: An exact mass-inflation solution}",
    journal = "Phys. Rev. Lett.",
    volume = "67",
    pages = "789--792",
    year = "1991",
    doi = "10.1103/PhysRevLett.67.789"
}

@article{Ovalle:2023vvu,
    author = "Ovalle, Jorge",
    title = "{Black holes without Cauchy horizons and integrable singularities}",
    eprint = "2305.00030",
    archivePrefix = "arXiv",
    primaryClass = "gr-qc",
    doi = "10.1103/PhysRevD.107.104005",
    journal = "Phys. Rev. D",
    volume = "107",
    number = "10",
    pages = "104005",
    year = "2023"
}

@article{Penrose:1969pc,
    author = "Penrose, R.",
    title = "{Gravitational collapse: The role of general relativity}",
    doi = "10.1023/A:1016578408204",
    journal = "Riv. Nuovo Cim.",
    volume = "1",
    pages = "252--276",
    year = "1969"
}

@article{Poisson:1989zz,
    author = "Poisson, Eric and Israel, W.",
    title = "{Inner-horizon instability and mass inflation in black holes}",
    doi = "10.1103/PhysRevLett.63.1663",
    journal = "Phys. Rev. Lett.",
    volume = "63",
    pages = "1663--1666",
    year = "1989"
}

@article{Poisson:1990eh,
    author = "Poisson, Eric and Israel, W.",
    title = "{Internal structure of black holes}",
    doi = "10.1103/PhysRevD.41.1796",
    journal = "Phys. Rev. D",
    volume = "41",
    pages = "1796--1809",
    year = "1990"
}

@article{Robinson:1975bv,
    author = "Robinson, D. C.",
    title = "{Uniqueness of the Kerr black hole}",
    doi = "10.1103/PhysRevLett.34.905",
    journal = "Phys. Rev. Lett.",
    volume = "34",
    pages = "905--906",
    year = "1975"
}

@article{Ruffini:1971bza,
    author = "Ruffini, Remo and Wheeler, John A.",
    title = "{Introducing the black hole}",
    doi = "10.1063/1.3022513",
    journal = "Phys. Today",
    volume = "24",
    number = "1",
    pages = "30",
    year = "1971"
}

@article{Silva:2025iip,
    author = "Silva, A. A. M. and Alencar, G. and Muniz, C. R. and Nilton, M. and Landim, R. R.",
    title = "{Topological thermodynamics of black holes: Revisiting the methods of winding numbers calculation}",
    eprint = "2511.06579",
    archivePrefix = "arXiv",
    primaryClass = "gr-qc",
    doi = "10.1016/j.physletb.2026.140589",
    journal = "Phys. Lett. B",
    volume = "879",
    pages = "140589",
    year = "2026"
}

@article{Smarr:1973zz,
    author = "Smarr, Larry",
    title = "{Mass Formula for Kerr Black Holes}",
    journal = "Phys. Rev. Lett.",
    volume = "30",
    pages = "71--73",
    year = "1973",
    note = "Erratum: Phys. Rev. Lett. 30, 521 (1973)",
    doi = "10.1103/PhysRevLett.30.71"
}

@article{Stelle:1977gr,
    author = "Stelle, K. S.",
    title = "{Classical Gravity with Higher Derivatives}",
    journal = "Gen. Rel. Grav.",
    volume = "9",
    pages = "353--371",
    doi = "10.1007/BF00760427",
    year = "1978"
}

@article{Stelle:1977ry,
    author = "Stelle, K. S.",
    title = "{Renormalization of Higher Derivative Quantum Gravity}",
    journal = "Phys. Rev. D",
    volume = "16",
    pages = "953--969",
    doi = "10.1103/PhysRevD.16.953",
    doi = "10.1103/PhysRevLett.114.171601",
    year = "1977"
}

@incollection{Tipler:1980,
    author = "Tipler, F. J. and Clarke, C. J. S. and Ellis, G. F. R.",
    title = "{Singularities and horizons: a review article}",
    booktitle = "General Relativity and Gravitation: One Hundred Years After the Birth of Albert Einstein",
    editor = "Held, A.",
    volume = "2",
    pages = "97--206",
    publisher = "Plenum",
    address = "New York",
    year = "1980"
}

@article{Visser:1992qh,
    author = "Visser, Matt",
    title = "{Dirty black holes: Thermodynamics and horizon structure}",
    eprint = "hep-th/9203057",
    archivePrefix = "arXiv",
    journal = "Phys. Rev. D",
    volume = "46",
    pages = "2445--2451",
    year = "1992",
    doi = "10.1103/PhysRevD.46.2445"
}

@article{Visser:1997tq,
    author = "Visser, Matt",
    title = "{General relativistic energy conditions: The Hubble expansion in the epoch of galaxy formation}",
    doi = "10.1103/PhysRevD.56.7578",
    journal = "Phys. Rev. D",
    volume = "56",
    pages = "7578--7587",
    year = "1997"
}

@article{Xiao:2023lap,
    author = "Xiao, Yong and Tian, Yu and Liu, Yu-Xiao",
    title = "{Extended Black Hole Thermodynamics from Extended Iyer-Wald Formalism}",
    journal = "Phys. Rev. Lett.",
    volume = "132",
    number = "2",
    pages = "021401",
    year = "2024",
    doi = "10.1103/PhysRevLett.132.021401"
}

@article{Wei:2022dzw,
    author = "Wei, Shao-Wen and Liu, Yu-Xiao and Mann, Robert B.",
    title = "{Black Hole Solutions as Topological Thermodynamic Defects}",
    eprint = "2208.01932",
    archivePrefix = "arXiv",
    primaryClass = "gr-qc",
    doi = "10.1103/PhysRevLett.129.191101",
    journal = "Phys. Rev. Lett.",
    volume = "129",
    number = "19",
    pages = "191101",
    year = "2022"
}

@article{Yang:2021civ,
    author = "Yang, Run-Qiu and Cai, Rong-Gen and Li, Li",
    title = "{Constraining the number of horizons with energy conditions}",
    eprint = "2104.03012",
    archivePrefix = "arXiv",
    primaryClass = "gr-qc",
    doi = "10.1088/1361-6382/ac4118",
    journal = "Class. Quant. Grav.",
    volume = "39",
    number = "3",
    pages = "035005",
    year = "2022"
}
\bibliographystyle{elsarticle-num}

\end{document}